\documentclass[aps,preprint,pdftex]{revtex4}
\usepackage{amssymb}
\usepackage{amsmath}
\usepackage{graphicx}
\usepackage{bm}
\usepackage{color}
\usepackage{caption}
\usepackage{epstopdf}
\usepackage[justification=raggedright]{caption}

\makeatletter

\newcommand{\Rmnum}[1]{\expandafter\@slowromancap\romannumeral #1@}
\makeatother

\begin{document}
\title{MHD stability analysis of ideal wall modes for CFETR upgrade phase-I scenario using NIMROD}

\author{
  Shikui Cheng$^1$, Ping Zhu$^{1,2,3}$,  Debabrata Banerjee$^1$, Xingting Yan$^1$, and CFETR Physics Team
 \\ $^1${\it CAS Key Laboratory of Geospace Environment and Department of Modern Physics \\
    University of Science and Technology of China, Hefei, Anhui 230026, PRC}
 \\ $^2${\it KTX Laboratory and Department of Modern Physics \\
    University of Science and Technology of China, Hefei, Anhui 230026, PRC}
 \\ $^3${\it Department of Engineering Physics \\
    University of Wisconsin-Madison,
    Madison, WI 53706, USA}
}

\email{pzhu@ustc.edu.cn}

\begin{abstract}
Ideal MHD stability of China Fusion Engineering Test Reactor (CFETR) upgrade phase-I baseline scenario has been evaluated using the initial value code NIMROD. The toroidal mode numbers for n=1-30 have been considered for stability analysis both in single-fluid and two-fluid MHD models. Our calculation rusults show that all modes are found to be unstable with characteristics of edge-localized modes. For n $\leq$ 13 modes, two-fluid MHD model gives a slightly higher growth rates than single-fluid MHD model, while for n $>$ 13 modes, this trend becomes opposite, which means two-fluid MHD model is needed for high-n mode analysis. In addition, $n=1-10$ modes are found to be more unstable with increasing wall position and eventually their growth rates approach values in the no-wall limit.
\end{abstract}

\maketitle

%===============================
% introduction
%===============================

\section{Introduction}

%=================
% paragraph 1
%=================

The China Fusion Engineering
Test Reactor (CFETR) project has been proposed for conducting
 experiments in plasma regimes of the future fusion reactor~\cite{Song14,Chan15}. The dual purpose
of performing long duration steady state operation with conservative physics parameters as well as demonstrating high end
fusion power gain has led to the design of the latest scenario larger in dimension than that of ITER. This project
is envisioned to resolve many advanced issues such as DEMO blanket and divertor solution, advancement of remote-handling facilities
for maintaining in-vessel components, performance of high annual duty factor of $0.3-0.5$, demonstration of tritium self-sufficiency with
target tritium breeding ratio greater than one. The upgraded CFETR design has two phases having same geometrical parameters: the phase-I is designed to
have more restricted stable parameter regimes with a target to yield less fusion power ($ \sim 200 MW$) , whereas phase-II is more reactor-like
to demonstrate high fusion power $> 1$ GW with gain $Q>15$.

%=================
% paragraph 2
%=================

Due to the planned requirements of high
beta and high value of non-inductive bootstrap current fraction in CFETR, both pressure driven and current driven modes are likely to be excited. The requirement of moderate (Phase-I)
to high fusion power gain (phase-II) in CFETR, would require higher pedestal top pressure value resulting in a steeper gradient in pressure profile near to last closed surface. The aim to
operate CFETR in nearly fully non-inductive regime, has proposed requirement of $50\%$ of bootstrap current at phase-I and $75\%$ at phase-II. These essential
requirements are expected to lead to the excitation of ideal MHD edge localized
peeling-ballooning modes or ELMs. For machines at future reactor scale, the sizes of ELMs are likely to be larger than the observations in currently operating medium
sized tokamaks like EAST~\cite{Li13}. The repetitive expulsion of stored plasma energy and particles outside of magnetic confined domain would lead to continuous degradation 
of fusion power and high damaging heat loads onto divertor and first wall components. 

%=================
% paragraph 3
%=================

This article reports the results of analysis of ideal modes for CFETR phase-I scenario  using both single-fluid and two-fluid models implemented in the extended MHD initial value code NIMROD~\cite{Sovinec04}. This baseline case is found to be unstable for
edge localized modes with toroidal mode numbers $n=1-30$. In addition, we have studied the effects of wall position. All n-modes are found to be less unstable when the wall gets closer to the plasma boundary, and they will be well stabilized inside a particular wall position.

%=================
% paragraph 4
%=================

The rest of this paper is organized as follows. Section~\ref{equilibrium} introduces the CFETR phase-I baseline equilibrium.
Section~\ref{equation} describes MHD model in NIMROD with the MHD equations considered.
In Section~\ref{result}, we present the numerical results in details.
Finally, the main points are summarized and conclusion is drawn in Section~\ref{summary}.

%============================
% equilibrium
%============================

\section{CFETR Upgrade Phase-I Equilibrium}
\label{equilibrium}
Both 0-D and 1.5D transport simulation methods have led to the latest phase of design.
In 2014-15, different 0-D system codes were employed to provide initial phases of design with relatively smaller size plasma and more conservative target
of fusion power as a starting point~\cite{Wan14,Chan15}. Later, more advanced scenarios have been designed for more optimized parameters including plasma size, normalized
beta, projected fusion power gain and bootstrap power drive fraction~\cite{Shi16}.
Besides 0-D calculation, 1.5D integrated modeling has been used to explore these scenarios as described in recent article~\cite{Jian17}.
 Now the immediate issue to address is whether these equilibrium profiles are stable or not in terms of the most dominant ideal and non-ideal MHD modes.

\begin{figure}[ht]
\centering
\includegraphics[width=0.45\textwidth,height=0.50\textheight]{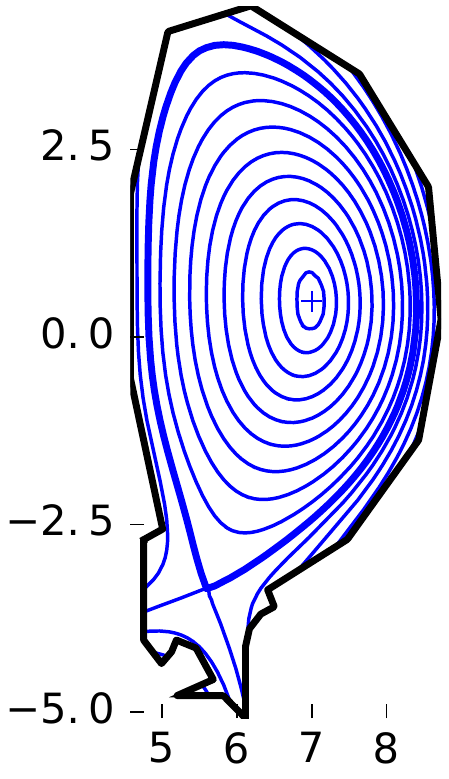}
\caption{Cross section of CFETR baseline design with currently proposed wall configuration.}
\label{fig:cfetr1015}
\end{figure}

\begin{figure}[ht]
\begin{minipage}{0.49\textwidth}
\includegraphics[width=1.0\textwidth,height=0.3\textheight]{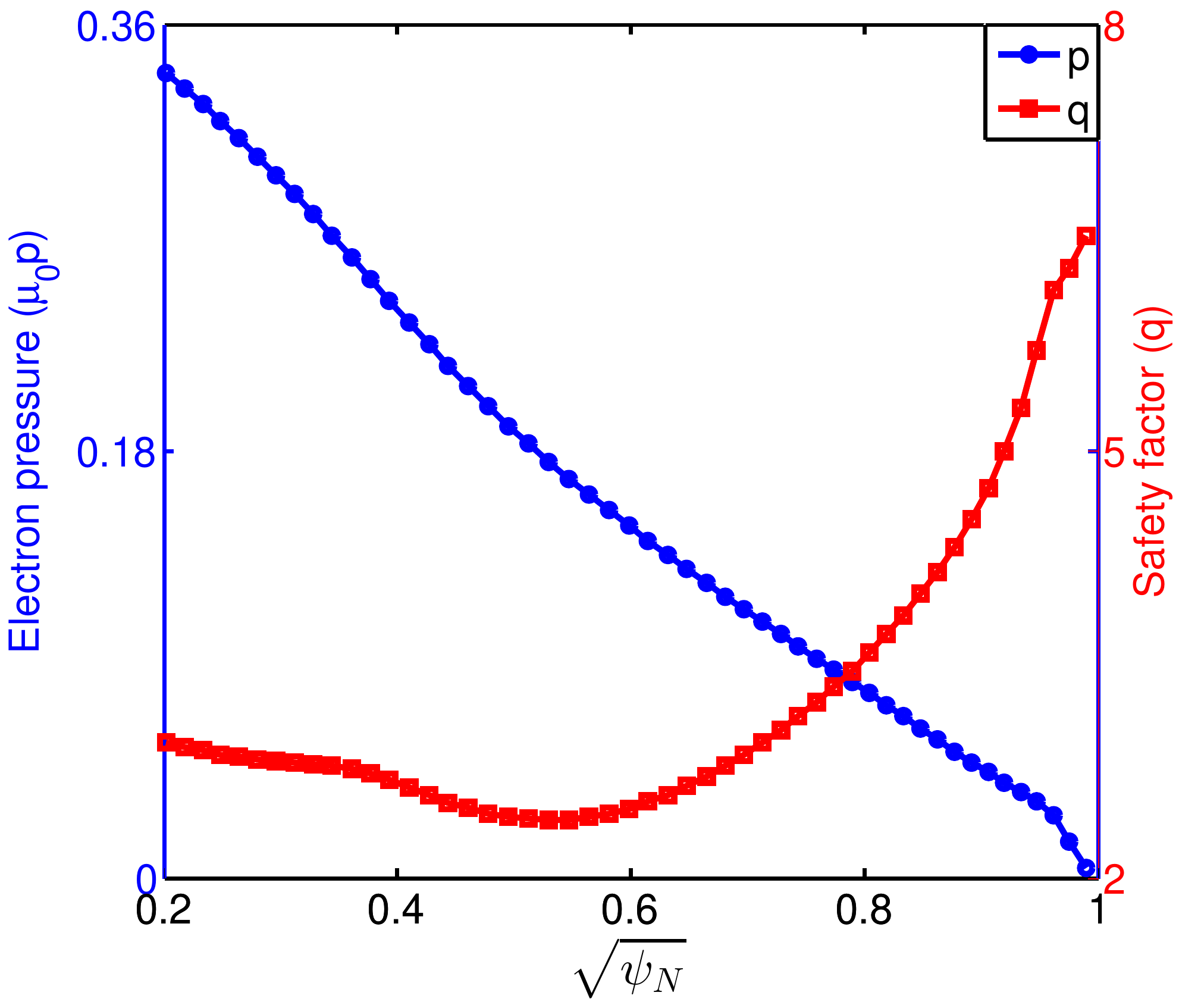}
\put(-220,180){\textbf{(a)}}
\end{minipage}
\begin{minipage}{0.495\textwidth}
\includegraphics[width=1.0\textwidth,height=0.31\textheight]{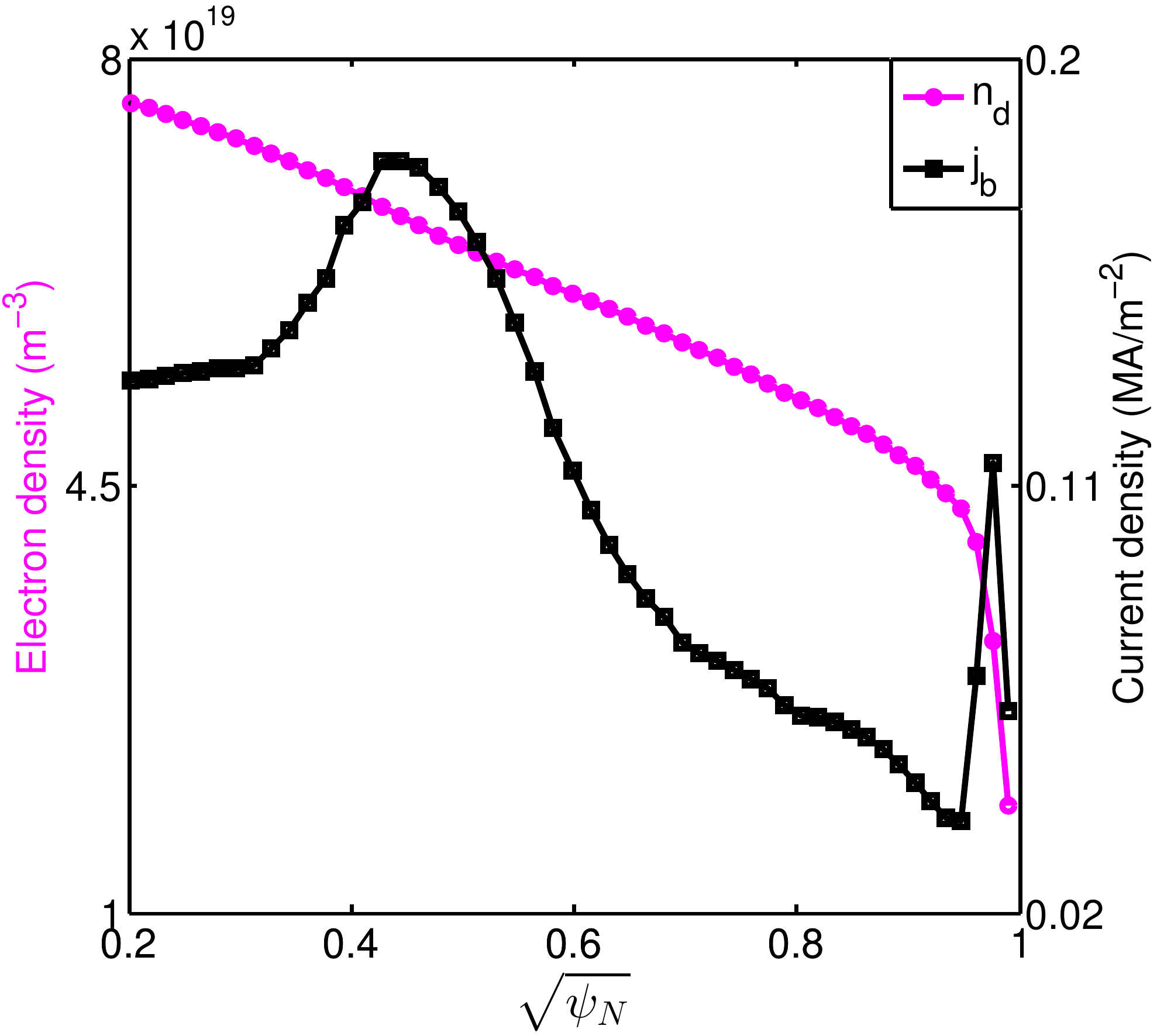}
\put(-220,180){\textbf{(b)}}
\end{minipage}
\caption{CFETR baseline phase I equilibrium profiles as functions of the root of normalized magnetic flux $\sqrt{\psi_N}$ for (a) pressure (blue line) and safety factor (red line); (b) current density (purple line) and toroidal rotation frequency (dark line).}
\label{fig:equilibrium}
\end{figure}

We consider the CFETR upgrade phase-I equilibrium with major radius 6.6m and minor radius 1.8m, as shown in Fig.\ref{fig:cfetr1015}. This equilibrium has self-consistently been generated through transport modelling in the OMFIT framework~\cite{Meneghini15,Meneghini16} using the auxiliary heating schemes such as neutral beam injection (NBI) and electron cyclotron wave (ECH, ECCD). As phase-I case is not designed for high fusion gain, the normalized $\beta_N$ is set to be $1.8$, which is meant to ensure this equilibrium away from stability limits, e.g. below than no wall beta limit of
$\beta_N \sim 4 \times l_i$, where $l_i$ is the plasma inductance. However, steep pressure gradient and high bootstrap current fraction at the edge pedestal may lead to the
excitation of ELMs (Fig.\ref{fig:equilibrium}).

%=================================
% model equations
%=================================
\section{Extended MHD Model in NIMROD}
\label{equation}
We use the NIMROD code~\cite{Sovinec04} for our stability analysis.
The MHD equations used in our NIMROD calculations are:
\begin{equation}
\frac{\partial n}{\partial t} + \nabla \cdot \left( n {\bf u}\right) = 0
\end{equation}
\begin{equation}
m n \left( \frac{\partial}{\partial t} + {\bf u} \cdot \nabla \right) {\bf u} = {\bf J} \times {\bf B} - \nabla p - \nabla \cdot \bf \Pi
\end{equation}
\begin{equation}
\frac{3}{2} \left( \frac{\partial}{\partial t} + {\bf u}_{\alpha} \cdot \nabla \right) T_{\alpha} = -n T_{\alpha} \nabla \cdot {\bf u}_{\alpha} - \nabla \cdot {\bf q}_{\alpha}\quad\quad (\alpha=i,e)
\end{equation}
\begin{equation}
\frac{\partial {\bf B}}{\partial t} = - \nabla \times \left[ \eta {\bf J} - {\bf u}\times{\bf B} + \frac{1}{ne} \left({\bf J}\times{\bf B} - \nabla p_e \right)\right]
\end{equation}
\begin{equation}
\mu_0 {\bf J} = \nabla \times {\bf B}, ~~~~~~~~~~~~~~~ \nabla \cdot {\bf B} = 0
\end{equation}
where {\bf u} is the center-of-mass flow velocity, $n$ the particle density, $m$ the ion mass, $p$ the combined pressure of electron ($p_e$) and ion ($p_i$), $\eta$ the resistivity,
 and ${\bf \Pi}$ the ion stress tensor. The initial value NIMROD code has been broadly applied to studying different ideal and non-ideal MHD processes in both fusion and space plasmas~\cite{Burke10,King16,Zhu13}. 

 Unlike the true vacuum model (i.e. no particle or current) used in the ideal MHD eigenvalue codes such as ELITE and AEGIS, NIMROD uses a vacuum-like halo region to
 model free boundary modes. The halo region is specified as a region with a low temperature,
 low density plasma, in contrast to the high density, high temperature
 plasma in the core region~\cite{Burke10}. This modeling is more physically relevant in the sense
 that the region between the plasma separatrix and vacuum vessel usually consists
 of relatively cold  plasma.

The Spitzer resistivity model is used in our simulation. The resistivity $\eta_{\parallel}$ along the magnetic field takes the form~\cite{Kuritsyn06,Trintchouk03,Zarnstorff90}:
\begin{equation}
\eta_{\parallel}^{Spitzer}=\frac{\sqrt{2m_e}Z_{eff}e^2ln\Lambda}{12\pi^{2/3}\epsilon_0^2T_e^{3/2}}
\end{equation}
where $T_e$ is the electron temperature, $Z_{eff}$ is the effective ionic charge, $ln\Lambda=ln(T_e^{3/2}\sqrt{\pi}Ze^3n^{1/2})$ is the Coulomb logarithm.
The crossfield or transverse resistivity is approximately twice as large as the parallel resistivity, i.e.  $\eta_{\perp}^{Spitzer}=1.96\times \eta_{\parallel}^{Spitzer}$.

%============================

\section{Ideal MHD Stability Analysis Results}
\label{result}

Stability analysis of CFETR upgrade phase-I scenario equilibrium has been carried out using NIMROD code in the context of extended-MHD
model. First, linear stability of toroidal modes $n=1-30$ has been calculated based on single-fluid MHD and Spitzer resistivity models. Then 
the same calculation is extended including two-fluid and finite Larmor radius 
effects. The wall is considered to be perfectly conducting throughout this calculation and the influence of wall position on the stability 
of MHD modes has been investigated. The detailed results of single-fluid and two-fluid MHD stability analysis are presented in the following 
two subsections.

\subsection{Comparison of results from single-fluid and two-fluid MHD models}
\begin{figure}[ht]
\centering
\includegraphics[width=0.8\textwidth,height=0.45\textheight]{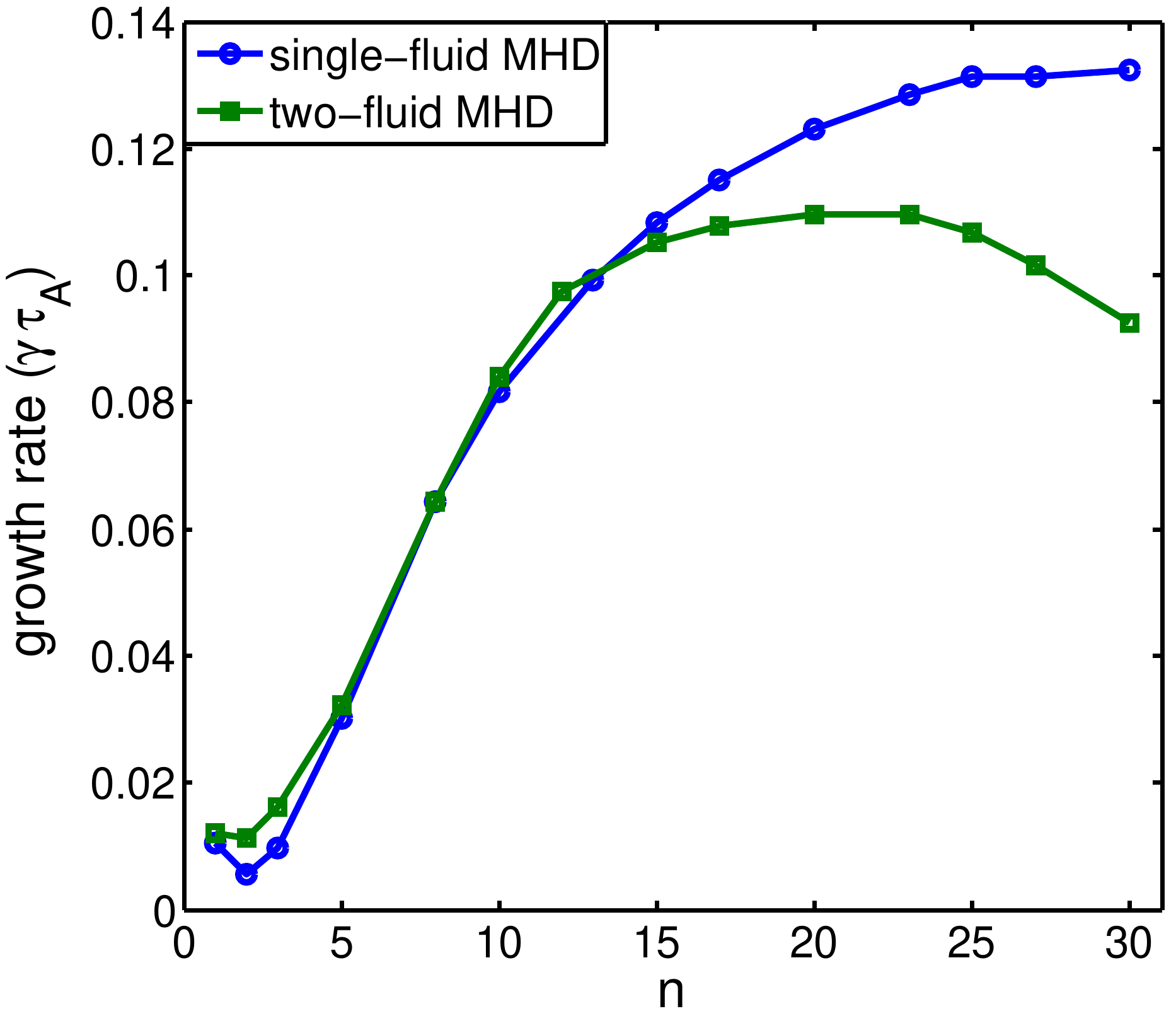}
\caption{Linear growth rates of edge localized modes as functions of toroidal mode number $n=1-30$ based on single-fluid MHD model (blue circle line) and  two-fluid MHD model (green square line).}
\label{fig:growth}
\end{figure}

Our NIMROD calculation of CFETR baseline equilibrium introduced in previous section, finds the excitation of peeling-ballooning modes localized in the edge pedestal
region. All the toroidal mode numbers ranging $n=1-30$ are found to have finite growth rate at the proposed CFETR ideal wall position at $b = 1.2a$,
where $b$ is the distance of wall from magnetic axis and $a$ the minor radius of plasma. However, no excitation of internal modes covering the core region of plasma is seen in the mode structure. 
Linear analysis shows the monotonous character of growth rate for n $\geq$ 2
modes in single-fluid MHD model as shown with blue line in
Fig.\ref{fig:growth}, while the growth rate of n = 1 mode is little higher than that of n=2 mode, where the growth rates are normalized with the Alfv\'enic time $\tau_A$.

For further study, we have employed two-fluid MHD model where two-fluid (i.e. Hall  and electron
diamagnetic effects) and finite Larmour radius effects are added to the
single-fluid MHD model in our calculation to
see the change in MHD growth rates and modes structure. The comparison of growth
rates of $n=1-30$ between results from single-fluid and two-fluid MHD are shown
in  Fig.~\ref{fig:growth}. The difference varies in different range of
toroidal mode numbers. 
Overall a slight increase in growth rate is noticed for the low-n modes with $n=1-5$ due to two-fluid effects, whereas the intermediate modes with $n=5-13$ have similar
growth rates in both models. A clear stabilizing role of two-fluid correction terms have been observed for the modes with $n>13$. The higher the toroidal mode number is, the stronger is the stabilizing effect from two-fluid MHD models.
For the mode with $n=30$, about $30.1\%$ reduction in growth rate is calculated.

%calculating growth rates are shown in  Fig.\ref{fig:growth}, comparing with the signle-fluid MHD results, one can find that at low n region (n $leq$ 13), two-fluid MHD model gives almost same growth rate as single-fluid MHD model, while for high-n modes (n $>$ 13), the results deviate, the non-monotonic nature of the two-fluid MHD growth rate curve shows that for toroidal mode number $n > 13$modes, we should include two-fluid effects.

\begin{figure}[ht]
\begin{minipage}{0.49\textwidth}
\includegraphics[width=1.0\textwidth,height=0.3\textheight]{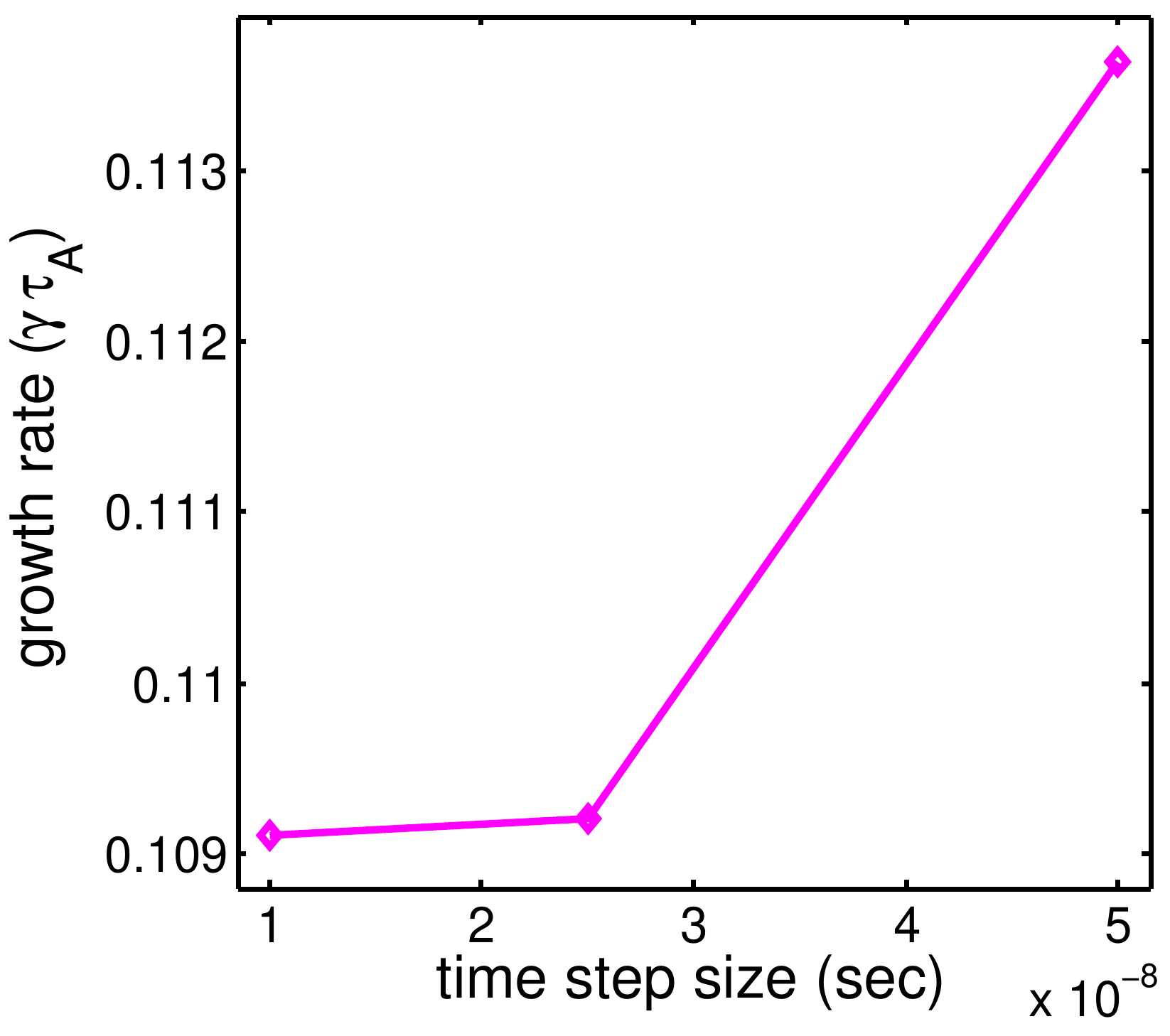}
\put(-220,180){\textbf{(a)}}
\end{minipage}
\begin{minipage}{0.49\textwidth}
\includegraphics[width=1.0\textwidth,height=0.3\textheight]{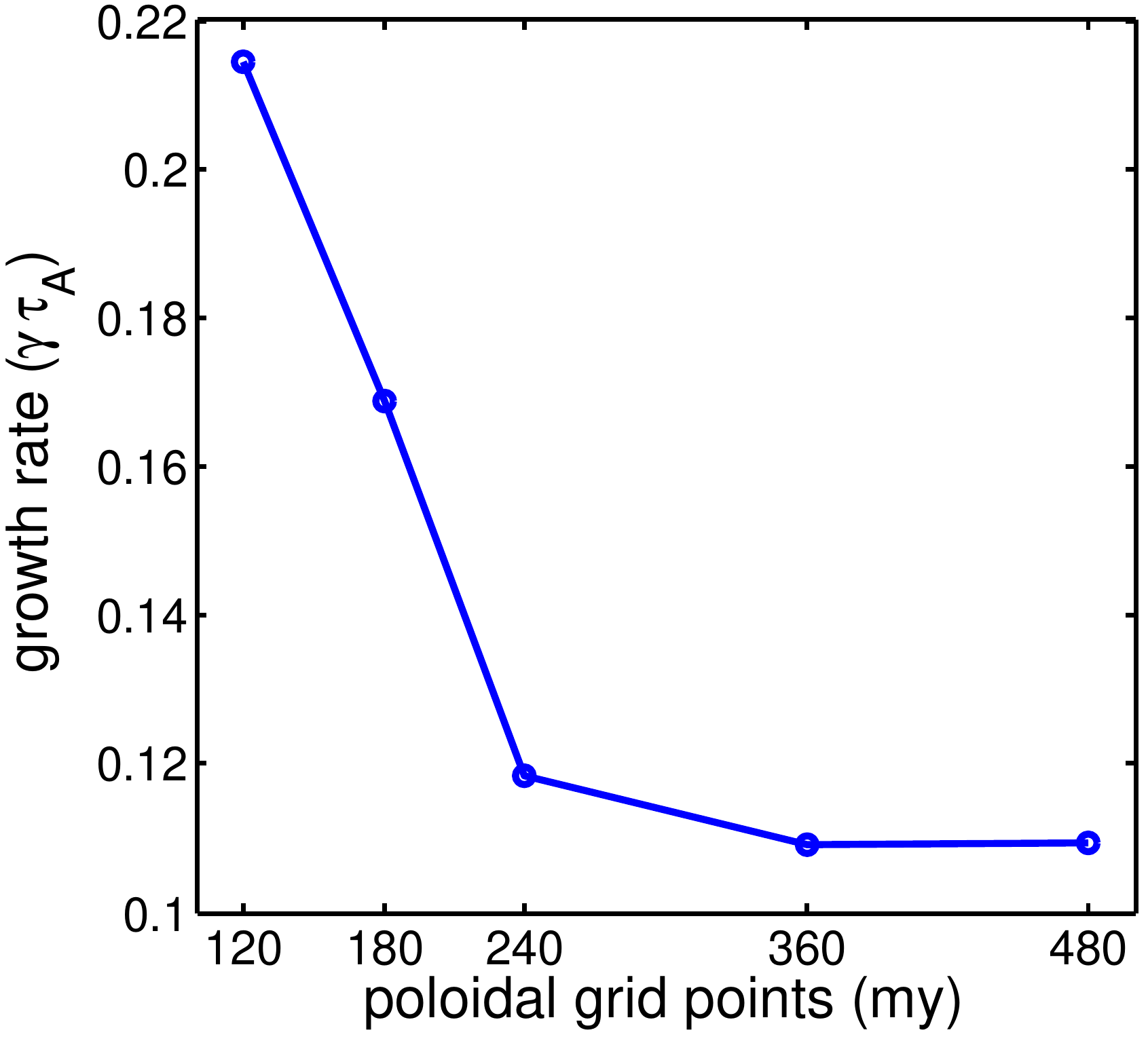}
\put(-220,180){\textbf{(b)}}
\end{minipage}

\begin{minipage}{0.49\textwidth}
\includegraphics[width=1.0\textwidth,height=0.3\textheight]{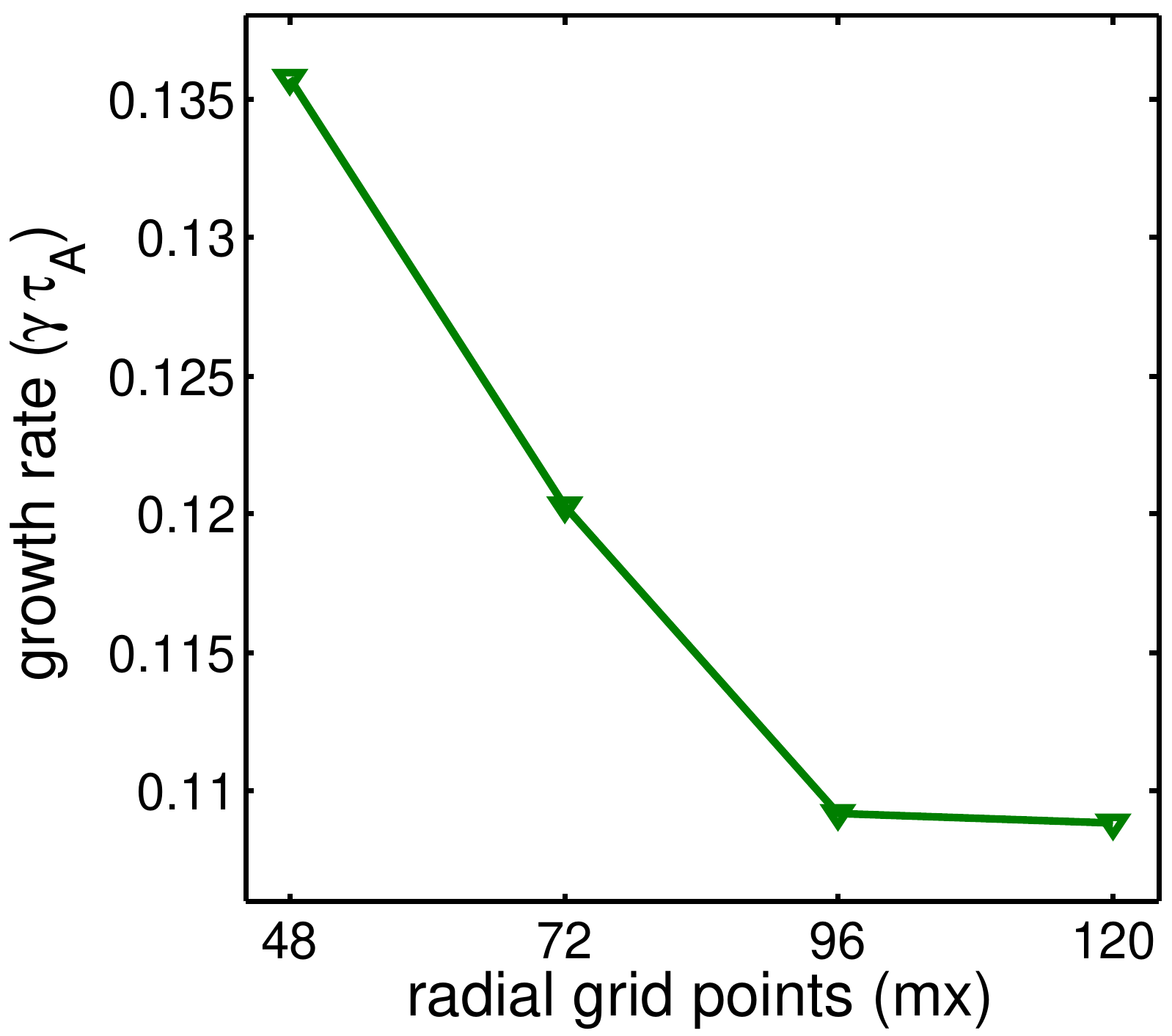}
\put(-220,180){\textbf{(c)}}
\end{minipage}
\begin{minipage}{0.49\textwidth}
\includegraphics[width=1.0\textwidth,height=0.3\textheight]{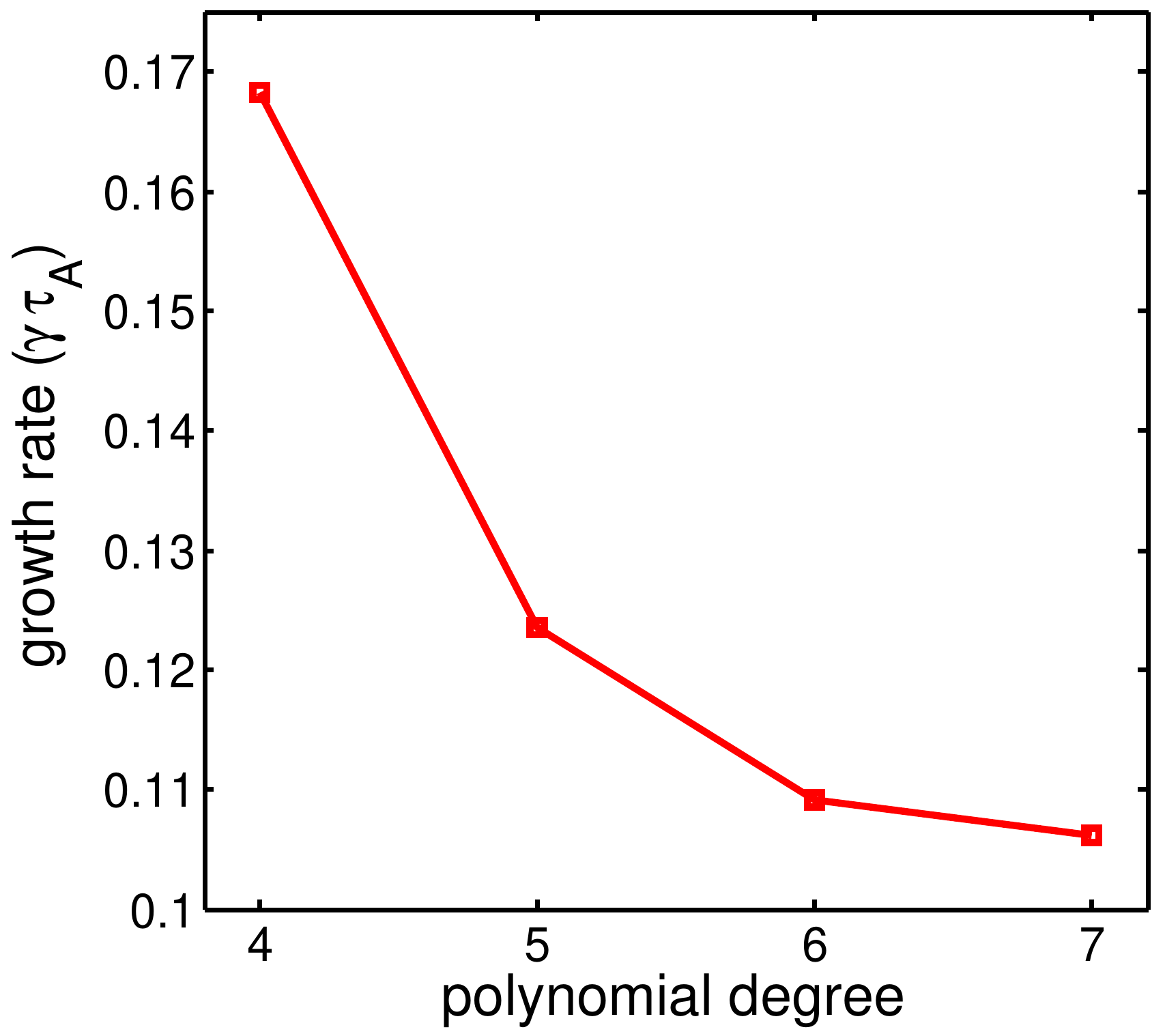}
\put(-220,180){\textbf{(d)}}
\end{minipage}
\caption{Linear growth rates of toroidal modes $n=20$ as functions of (a) time step size; (b) poloidal grid points; (c) radial grid points and (d) polynomial degree of finite element.}
\label{fig:convergence}
\end{figure}

Convergence test
has been carried out for time step size,  poloidal grid points, radial grid points and polynomial degree of finite element basis used in NIMROD calculation,  and the n=20 mode case is shown in Fig.~\ref{fig:convergence} for example. The growth rate of mode
 reaches convergence when the time step size decreases from $\Delta t=2.5\times 10^{-8}$ to $\Delta t=1.0\times 10^{-8}$, the poloidal grid number increases from 360 to 480, and the radial grid number increases from 96 to 120. Although there is small difference in growth between polynomial degree 6 and 7, the relative change (($\gamma_{poly=6}-\gamma_{ploy=7})/\gamma_{poly=6}$) in growth rate is about 2.7\% . These results indicate that the key numerical parameters used in our simulations are well within the converged regimes.

\begin{figure}[ht]
\begin{minipage}{0.47\textwidth}
\includegraphics[width=1.0\textwidth,height=0.35\textheight]{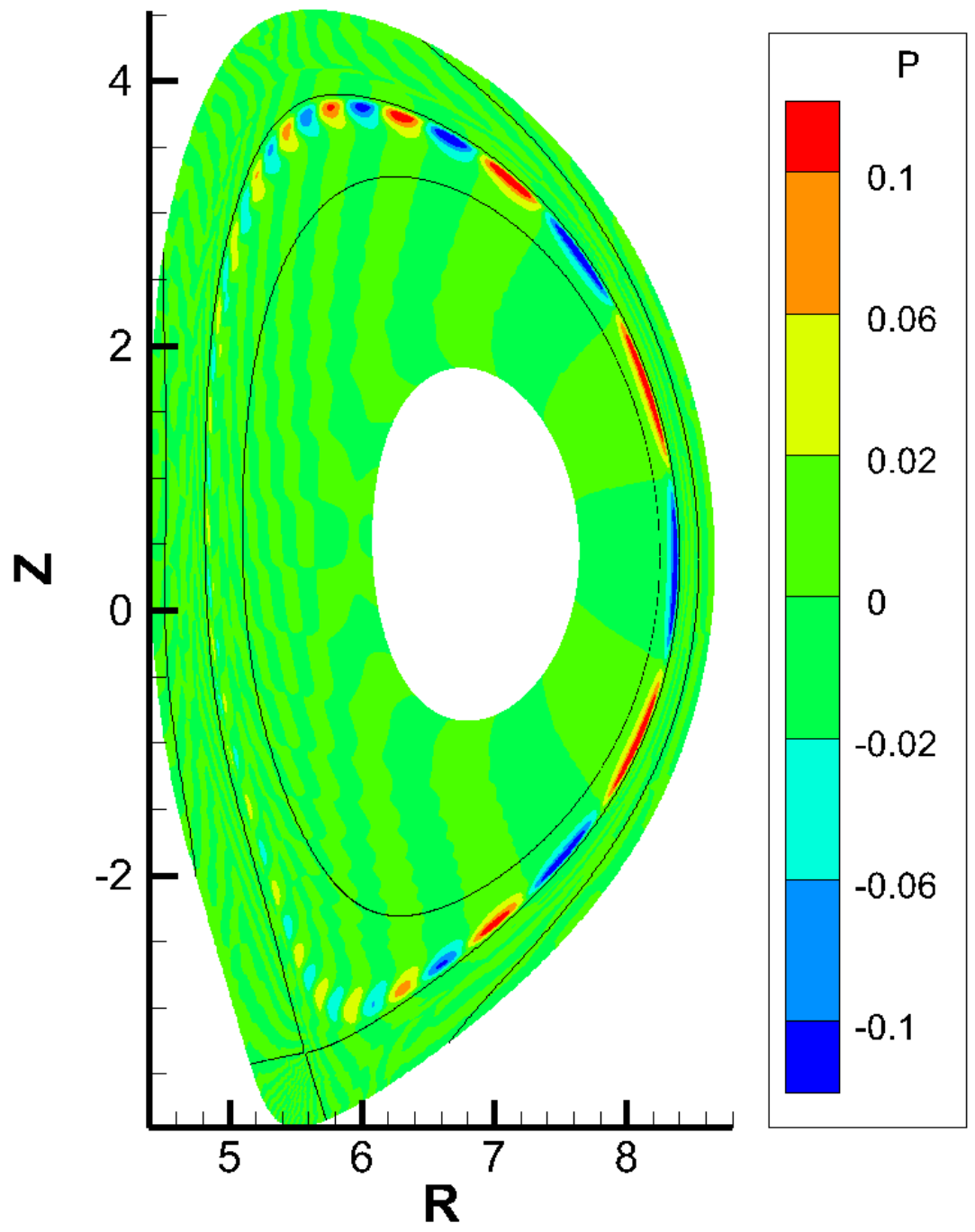}
\put(-220,200){\textbf{(a)}}
\put(-80,200){\textbf{n=3}}
\end{minipage}
\begin{minipage}{0.47\textwidth}
\includegraphics[width=1.0\textwidth,height=0.35\textheight]{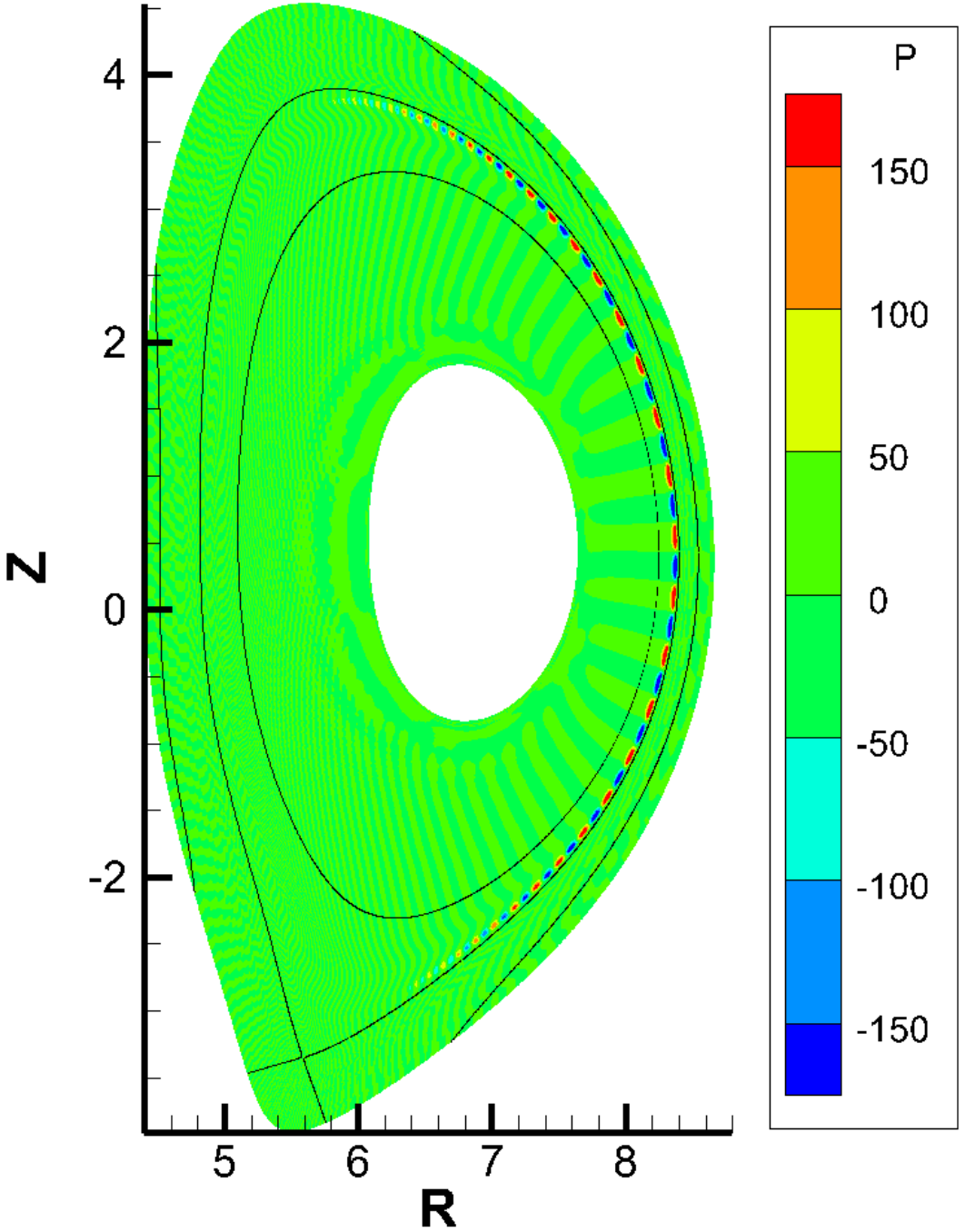}
\put(-220,200){\textbf{(b)}}
\put(-80,200){\textbf{n=20}}
\end{minipage}

\begin{minipage}{0.47\textwidth}
\includegraphics[width=1.0\textwidth,height=0.35\textheight]{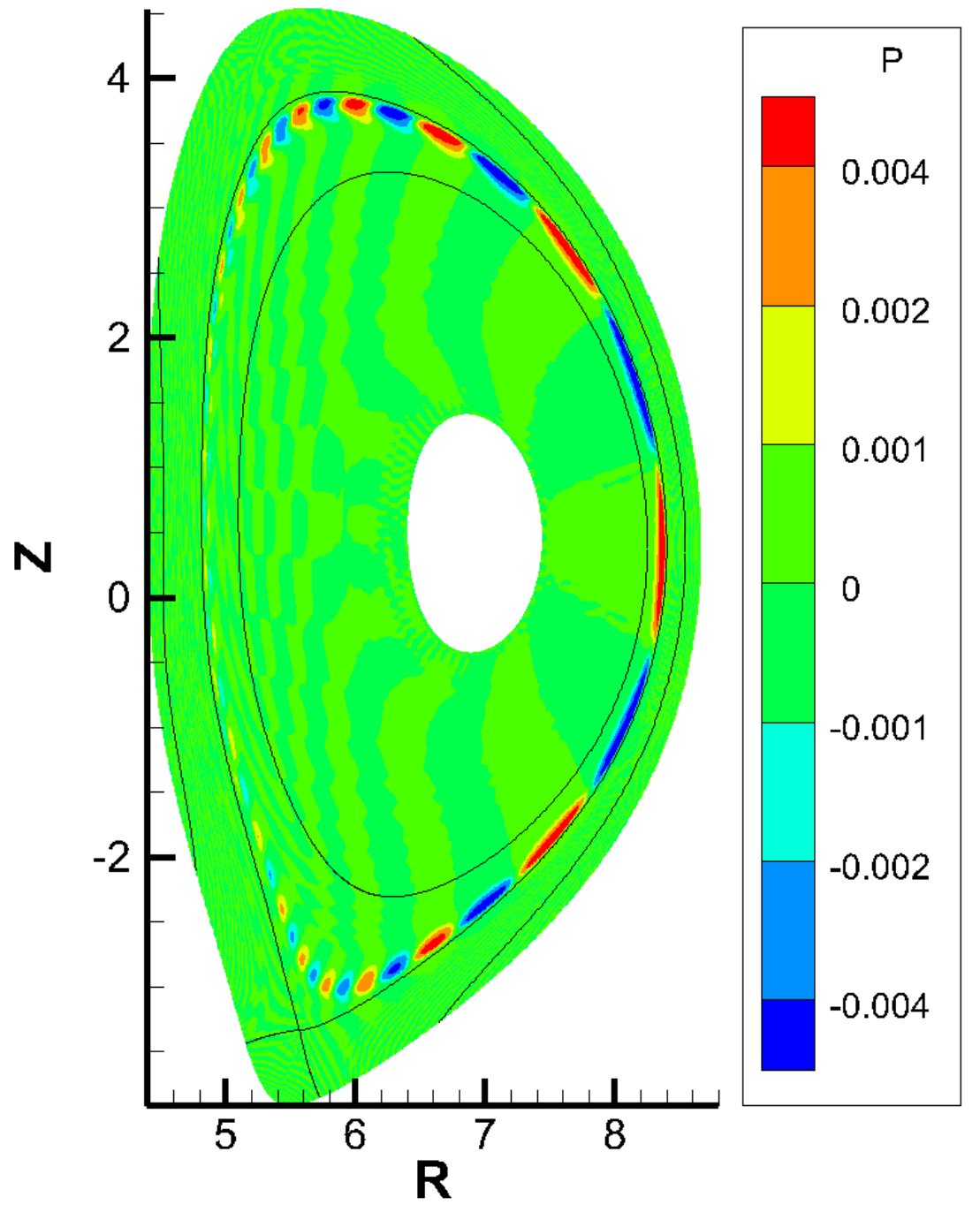}
\put(-220,180){\textbf{(c)}}
\end{minipage}
\begin{minipage}{0.47\textwidth}
\includegraphics[width=1.0\textwidth,height=0.35\textheight]{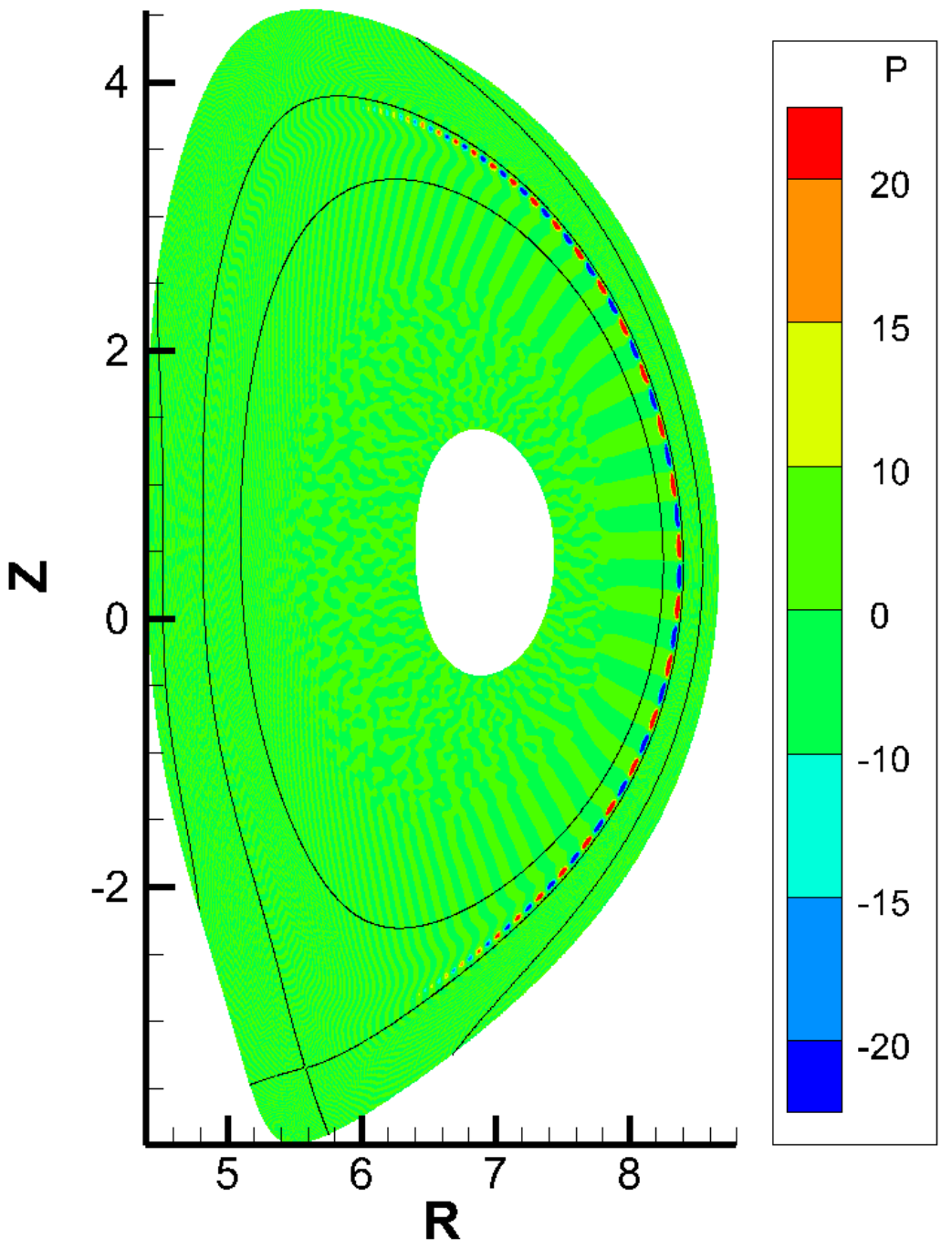}
\put(-220,180){\textbf{(d)}}
\end{minipage}
\caption{Colored contours represent pressure perturbation for (a) $n=3$ mode in single-fluid MHD model, (b) $n=20$ mode in single-fluid MHD model, (c) $n=3$ mode in two-fluid MHD model and (d) $n=20$ mode in two-fluid MHD model. Solid-line contour represents equilibrium magnetic flux function.}
\label{fig:contour1}
\end{figure}

\begin{figure}[ht]
\begin{minipage}{0.47\textwidth}
\includegraphics[width=1.0\textwidth,height=0.35\textheight]{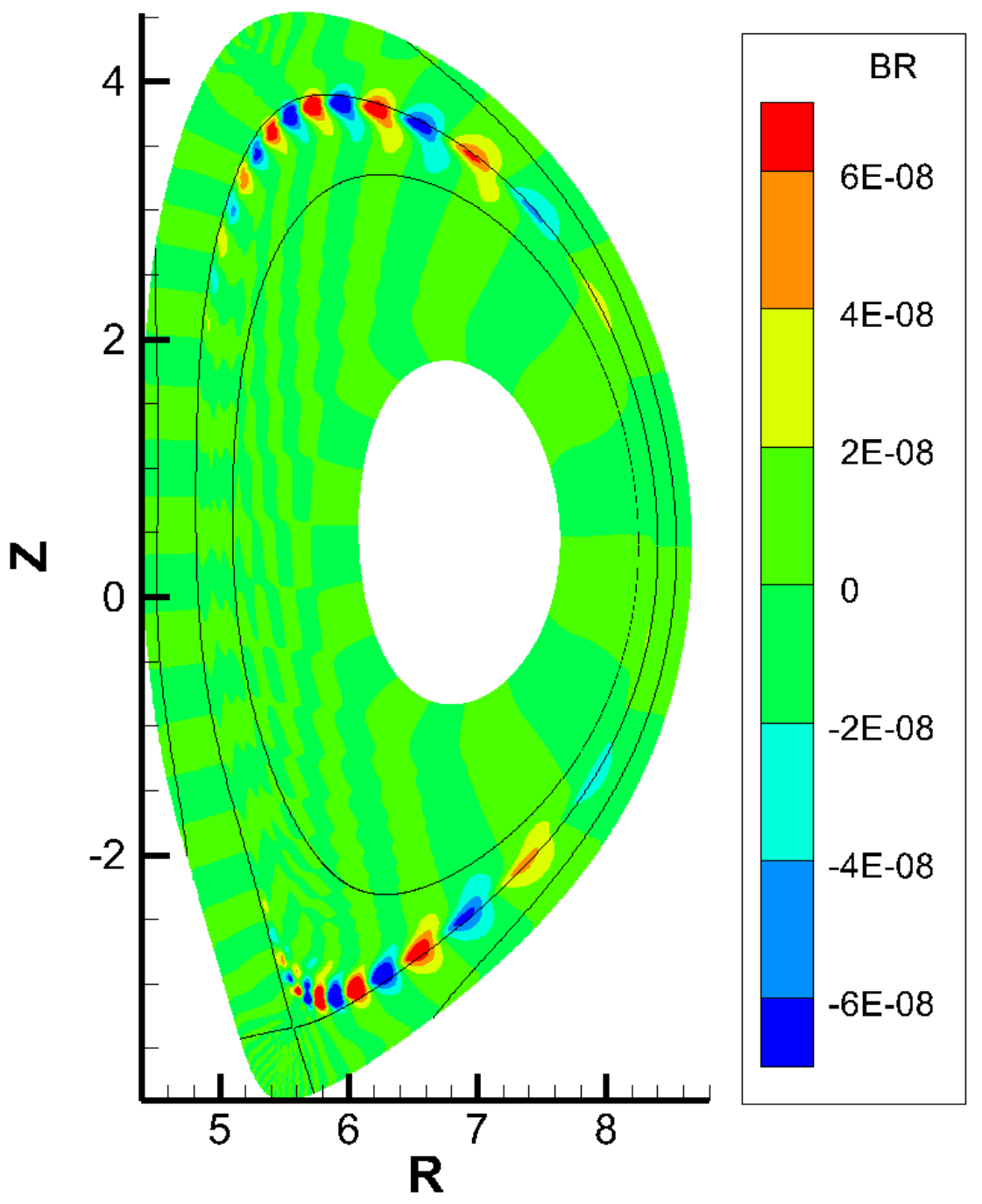}
\put(-220,180){\textbf{(a)}}
\end{minipage}
\begin{minipage}{0.47\textwidth}
\includegraphics[width=1.0\textwidth,height=0.35\textheight]{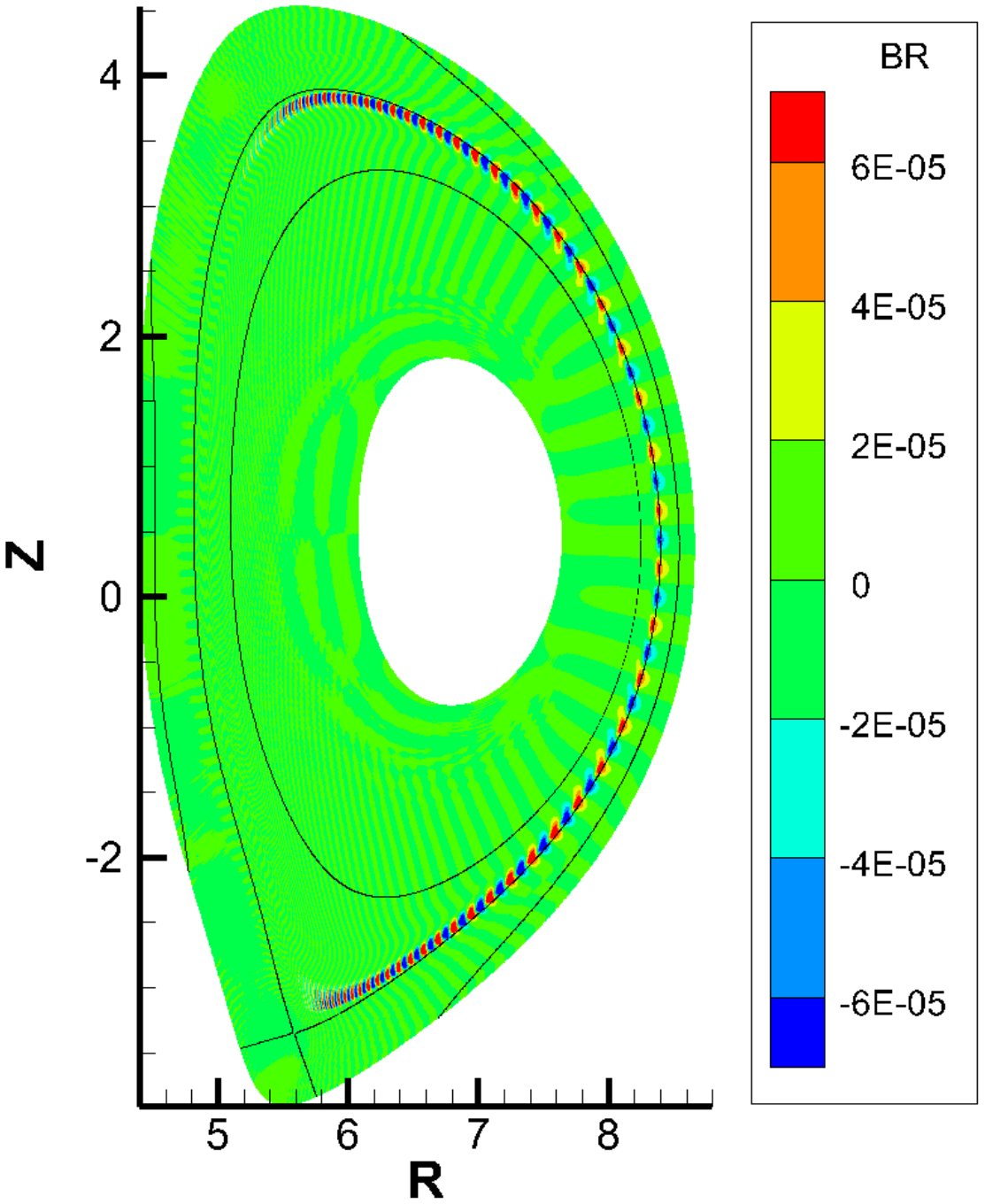}
\put(-220,180){\textbf{(b)}}
\end{minipage}

\begin{minipage}{0.47\textwidth}
\includegraphics[width=1.0\textwidth,height=0.35\textheight]{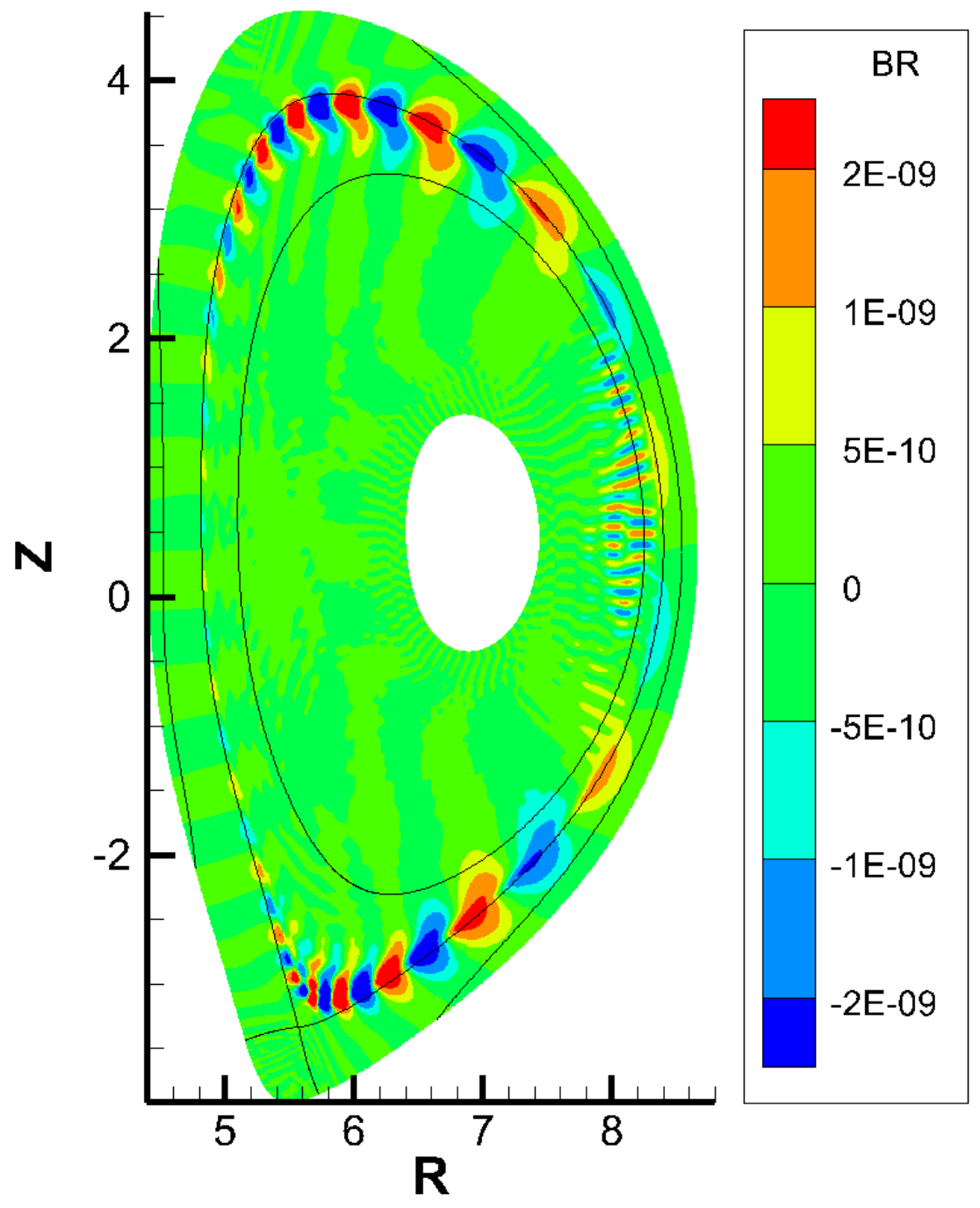}
\put(-220,180){\textbf{(c)}}
\end{minipage}
\begin{minipage}{0.47\textwidth}
\includegraphics[width=1.0\textwidth,height=0.35\textheight]{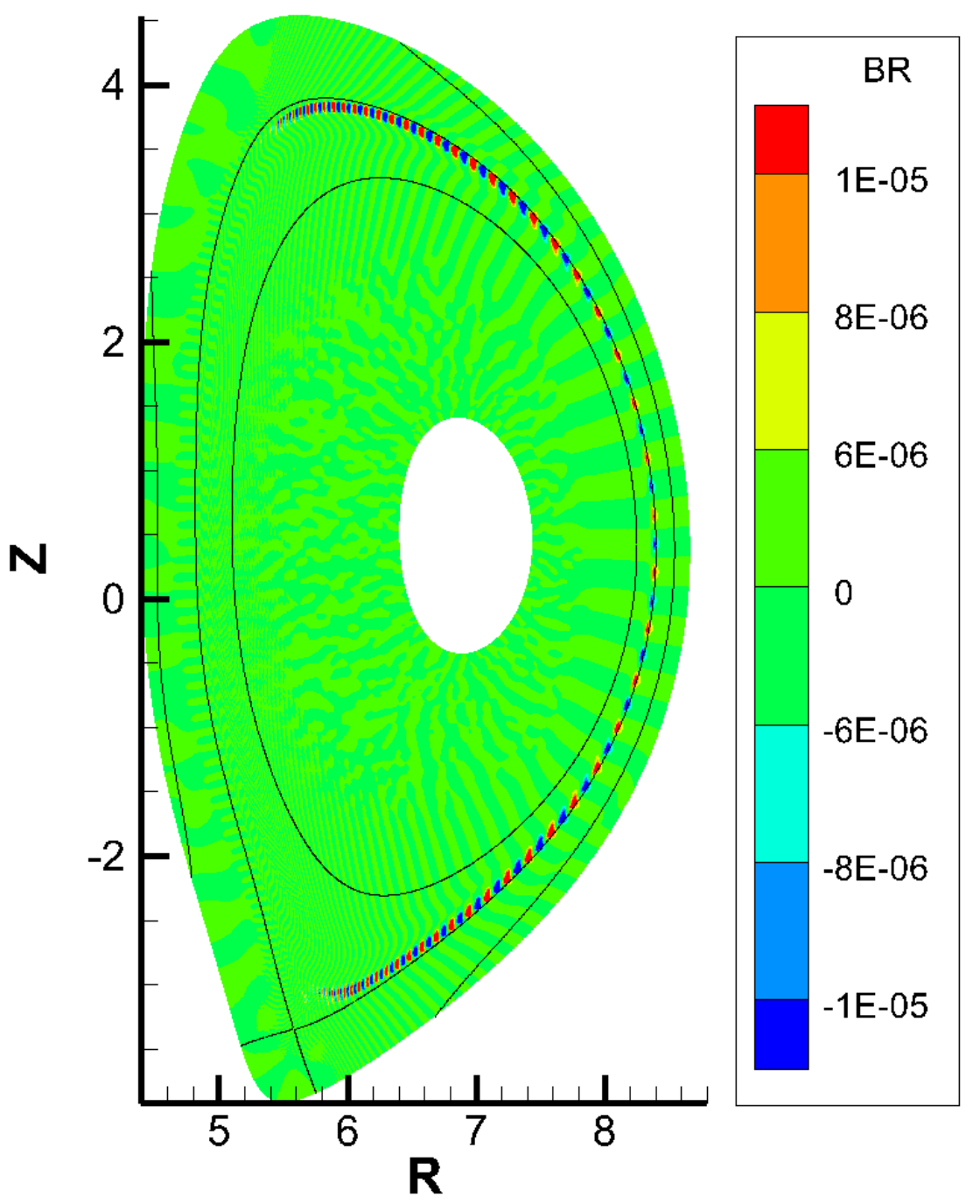}
\put(-220,180){\textbf{(d)}}
\end{minipage}
\caption{Colored contours represent radial magnetic perturbation in poloidal plane for (a) $n=3$ mode in single-fluid MHD model, (b) $n=20$ mode in single-fluid MHD model, (c) $n=3$ mode in two-fluid MHD model and (d) $n=20$ mode in two-fluid MHD model. Solid-line contour represents equilibrium magnetic flux function.}
\label{fig:contour2}
\end{figure}

The detailed structure of modes $n=3$ and $20$ for both single-fluid and two-fluid MHD models are generated. The perturbed pressure (Fig.\ref{fig:contour1}) and the radial component of magnetic field (Fig.\ref{fig:contour2}) are plotted in the poloidal plane,  where the dark contour lines of poloidal magnetic flux function in each plot show the locations of separatrices. All these modes are very close to the separatrix from inside in the pedestal region and show features of the peeling-ballooning mode structures.

\subsection{Wall stabilization effects}

\begin{figure}[ht]
\begin{minipage}{0.49\textwidth}
\includegraphics[width=1.0\textwidth,height=0.3\textheight]{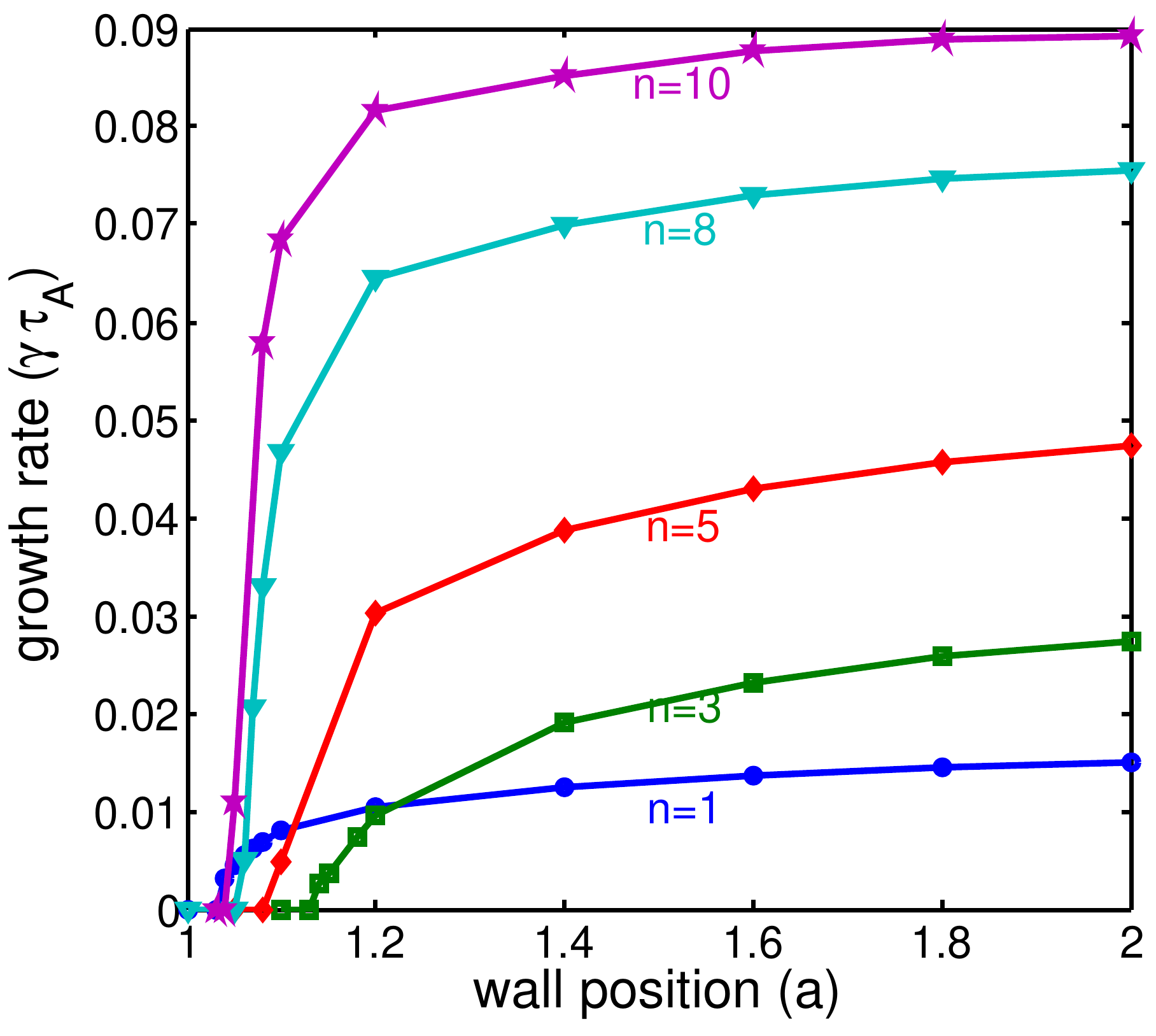}
\put(-220,180){\textbf{(a)}}
\end{minipage}
\begin{minipage}{0.495\textwidth}
\includegraphics[width=1.0\textwidth,height=0.31\textheight]{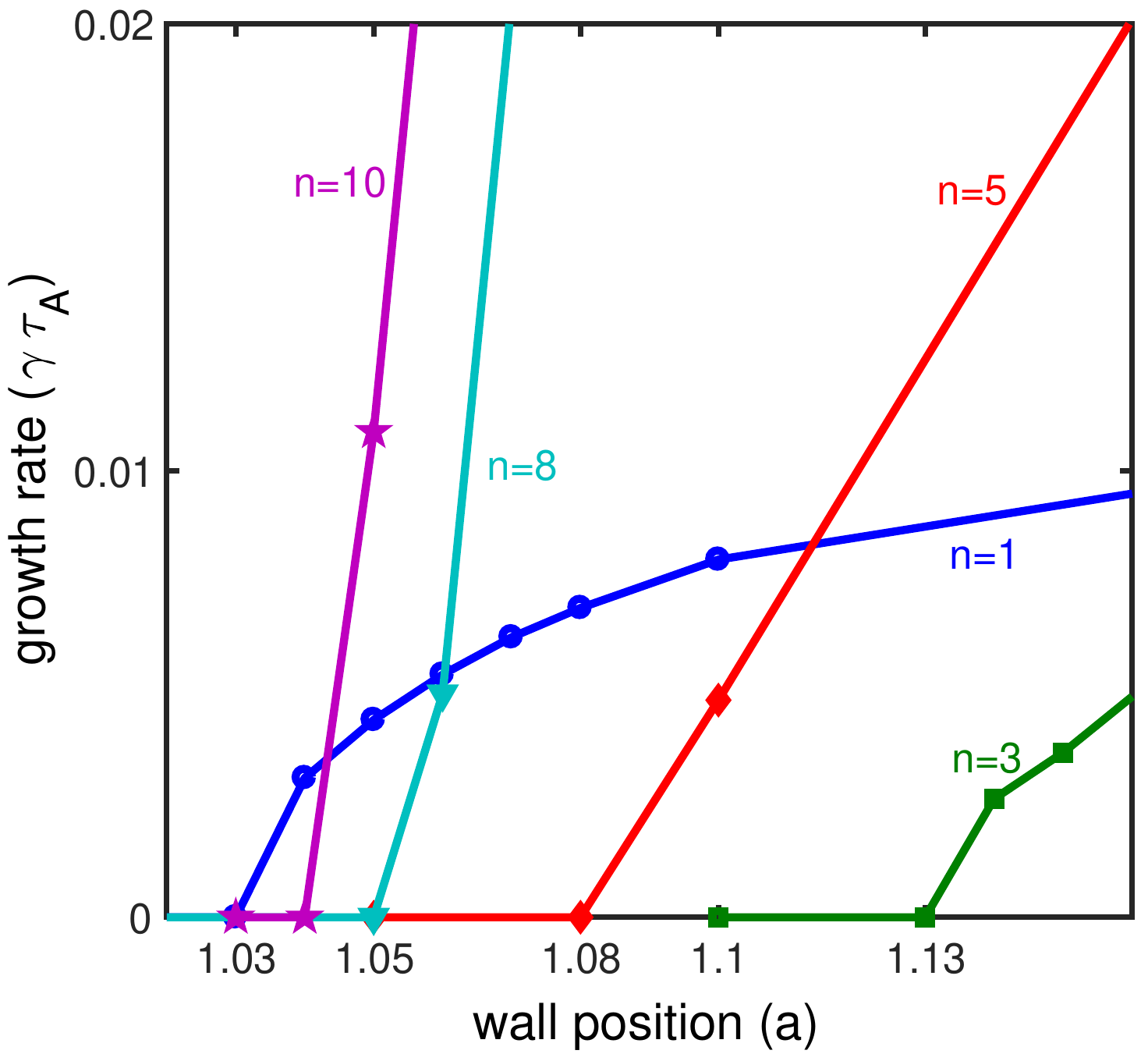}
\put(-220,180){\textbf{(b)}}
\end{minipage}
\caption{Linear growth rates of edge localized modes as functions of wall position for different toroidal mode number $n$ based on single-fluid MHD model.}
\label{fig:wall_growth}
\end{figure}

  To illustrate the wall position effects and provide physics base for the engineering design on the optimal choice of wall position of CFETR, we  calculate the growth rates of low-$n$ modes ($n=1, 3, 5, 8, 10$) with wall position varying from $b=1.0a$ to $2.0a$ sequentially. Single-fluid MHD model is used for calculation because two-fluid effects on low-$n$ modes are very weak. The wall is set to be ideal, fully conducting and conformal to the plasma edge shape in our calculation.
The main results are summarized in Fig.\ref{fig:wall_growth}.
 The growth rate initially increase as the wall position increases from $b=1.0a$ to $1.6a$, then becomes constant as $b>1.6a$ for the considered modes.
 
 As expected, these modes become less unstable when wall position gets closer to the plasma boundary. They become fully stabilized when wall position is within certain but different radius respectively.
 Specifically, $n=1$ mode is stabilized at $b=1.03a$, $n=3$ mode at $b=1.13a$, $n=5$ mode at $b=1.08a$, $n=8$ mode at $b=1.05a$ and $n=10$ mode at $b=1.04a$ for example.
It should be noted that in reality the wall is not perfectly conducting, it could bring in another essential instability such as resistive wall mode (RWM) \cite{Ward95}. 

% real wall

\begin{figure}[ht]
\centering
\includegraphics[width=0.8\textwidth,height=0.45\textheight]{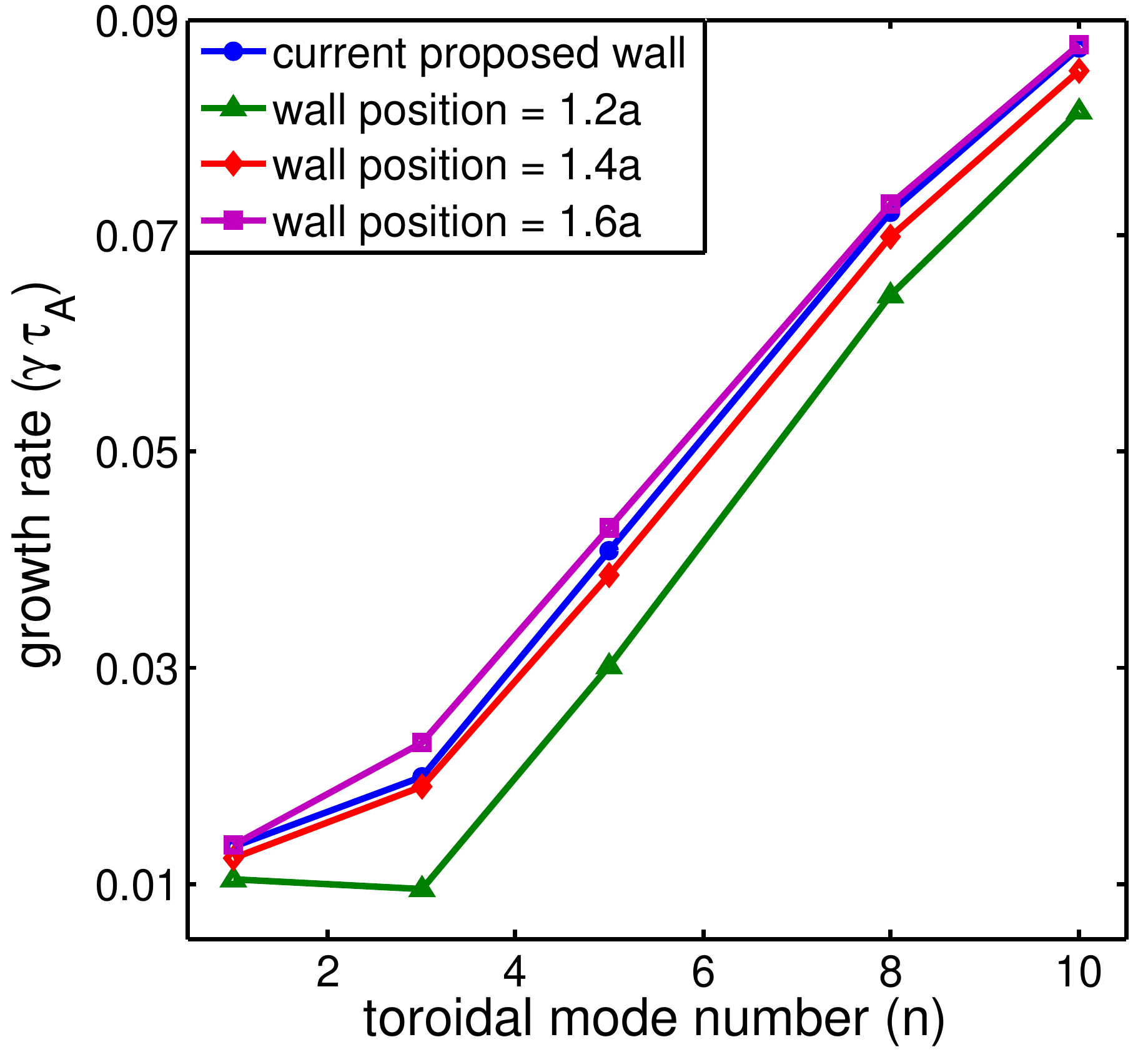}
\caption{Linear growth rates of edge localized modes as functions of different wall shape and position for toroidal mode number $n=1-10$ based on single-fluid MHD model.}
\label{fig:rw_growth}
\end{figure}

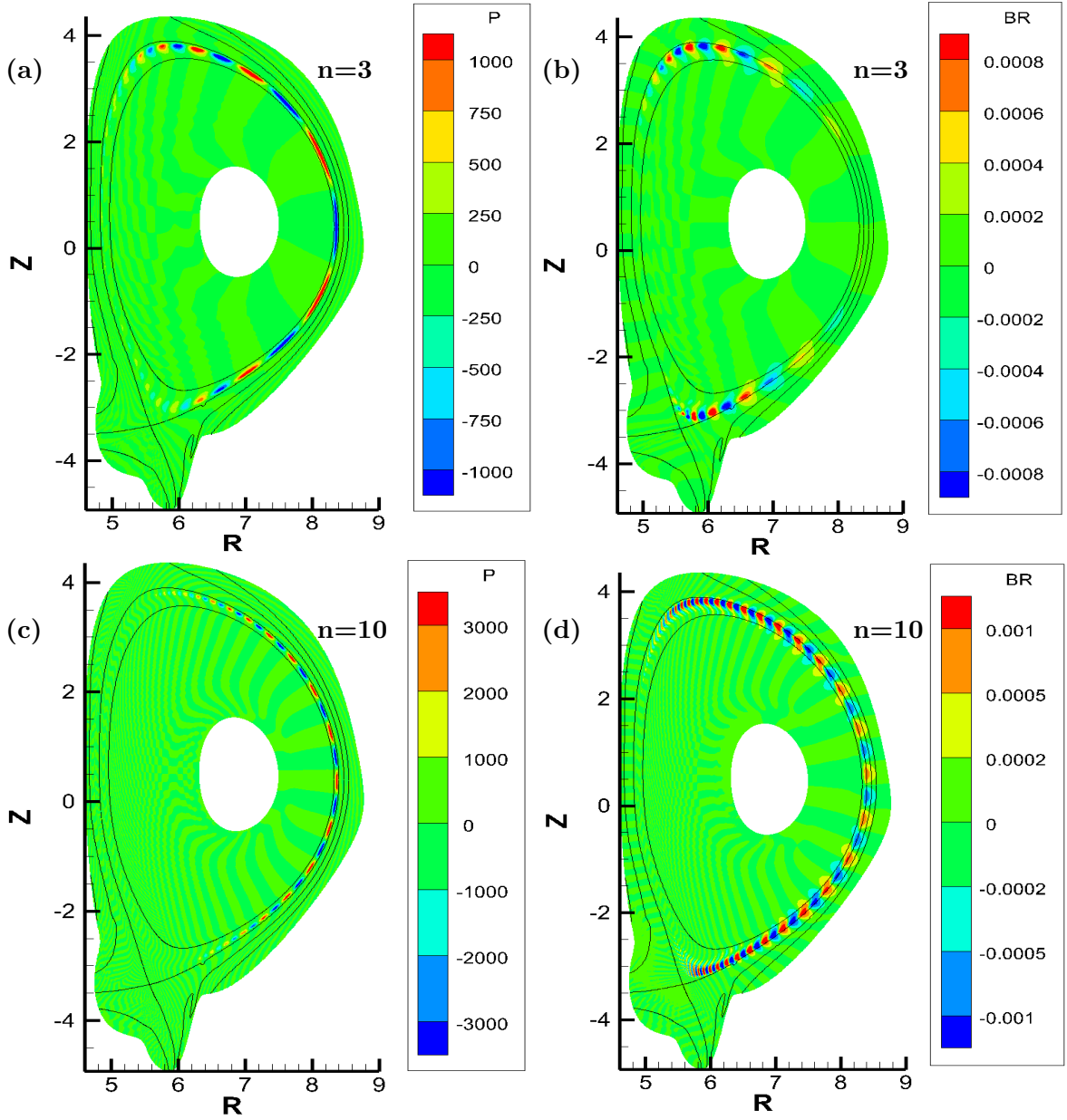
\begin{figure}[ht]
\begin{minipage}{0.47\textwidth}
\includegraphics[width=1.0\textwidth,height=0.35\textheight]{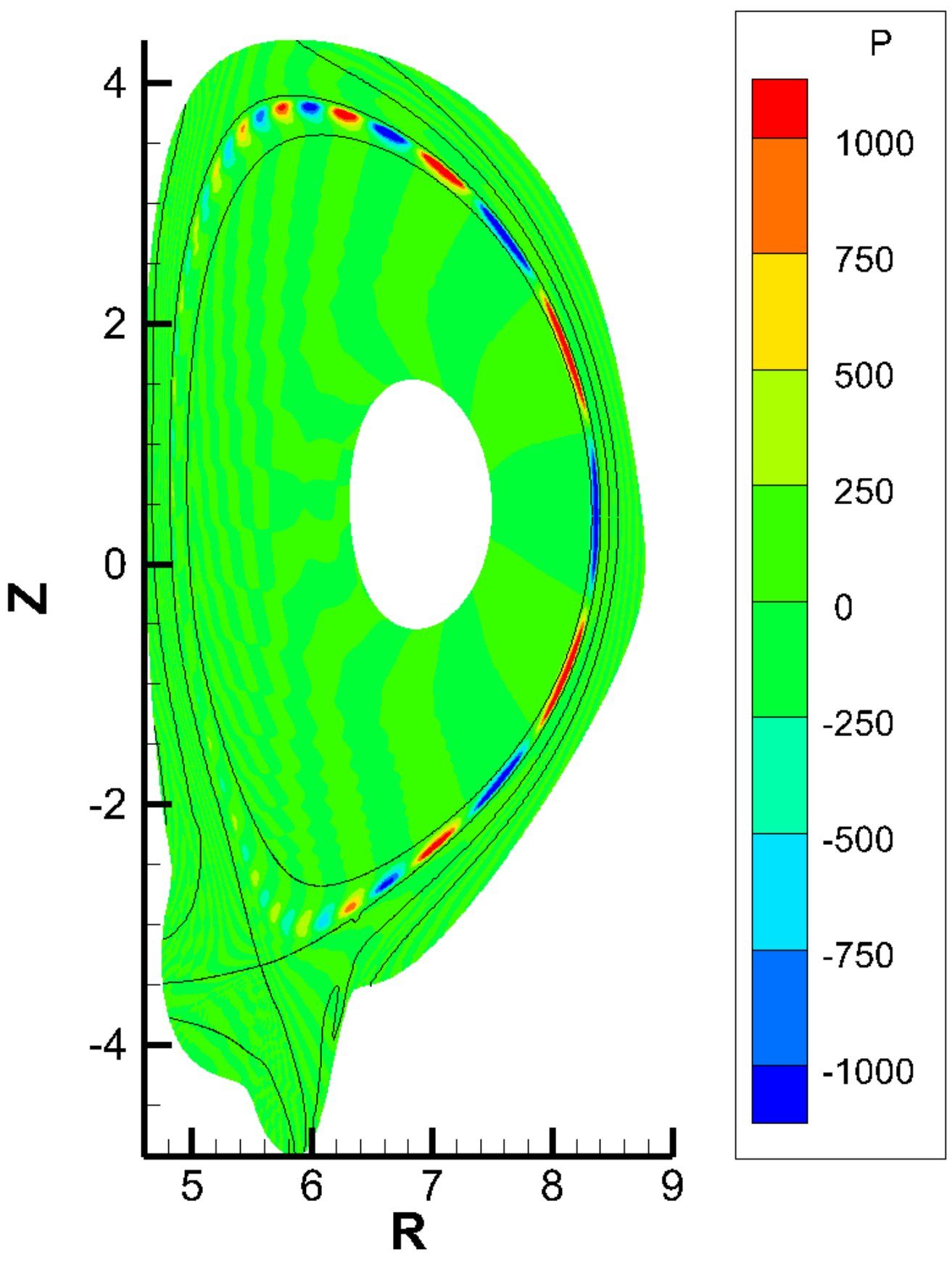}
\put(-220,200){\textbf{(a)}}
\put(-90,200){\textbf{n=3}}
\end{minipage}
\begin{minipage}{0.47\textwidth}
\includegraphics[width=1.0\textwidth,height=0.35\textheight]{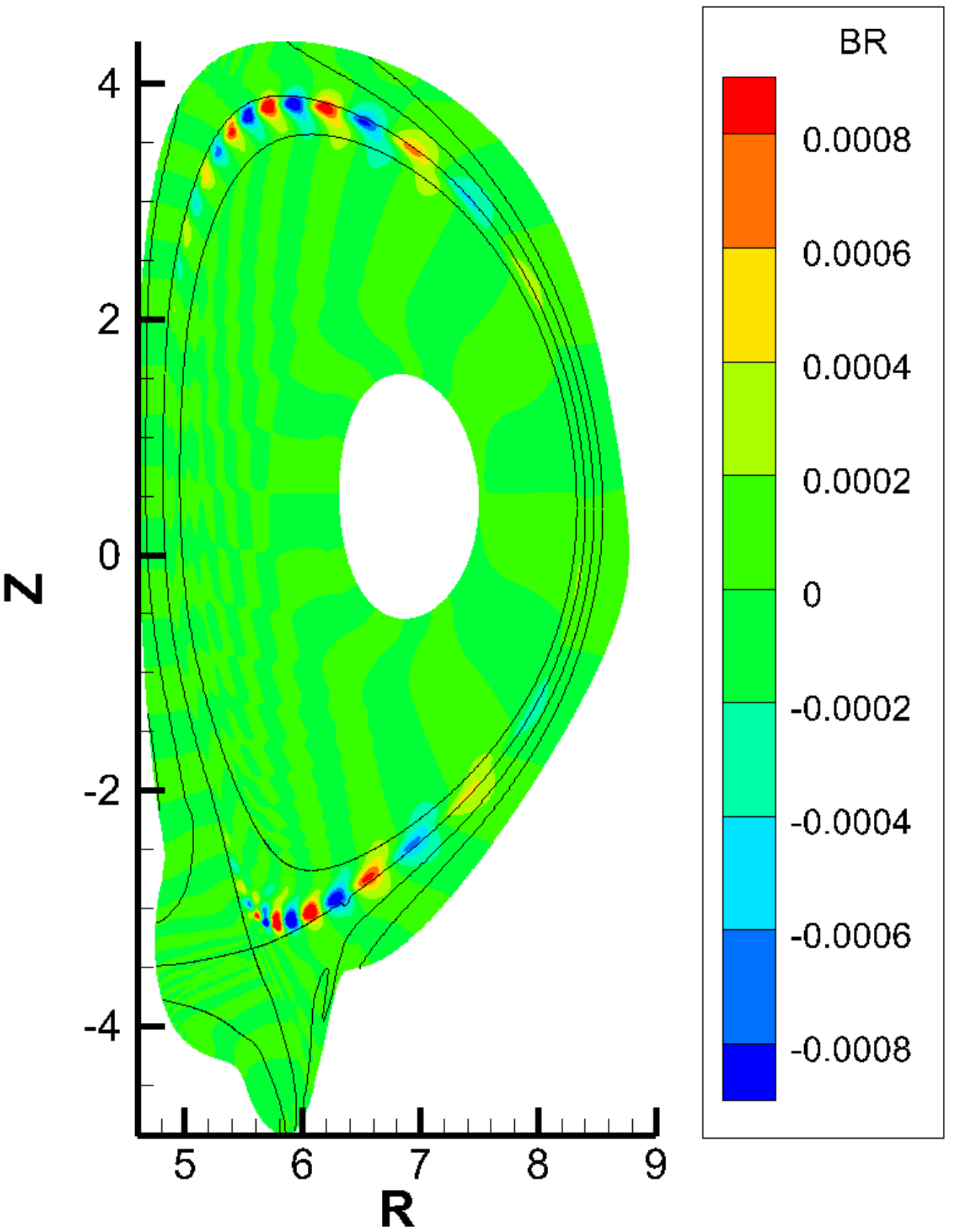}
\put(-220,200){\textbf{(b)}}
\put(-90,200){\textbf{n=3}}
\end{minipage}

\begin{minipage}{0.47\textwidth}
\includegraphics[width=1.0\textwidth,height=0.35\textheight]{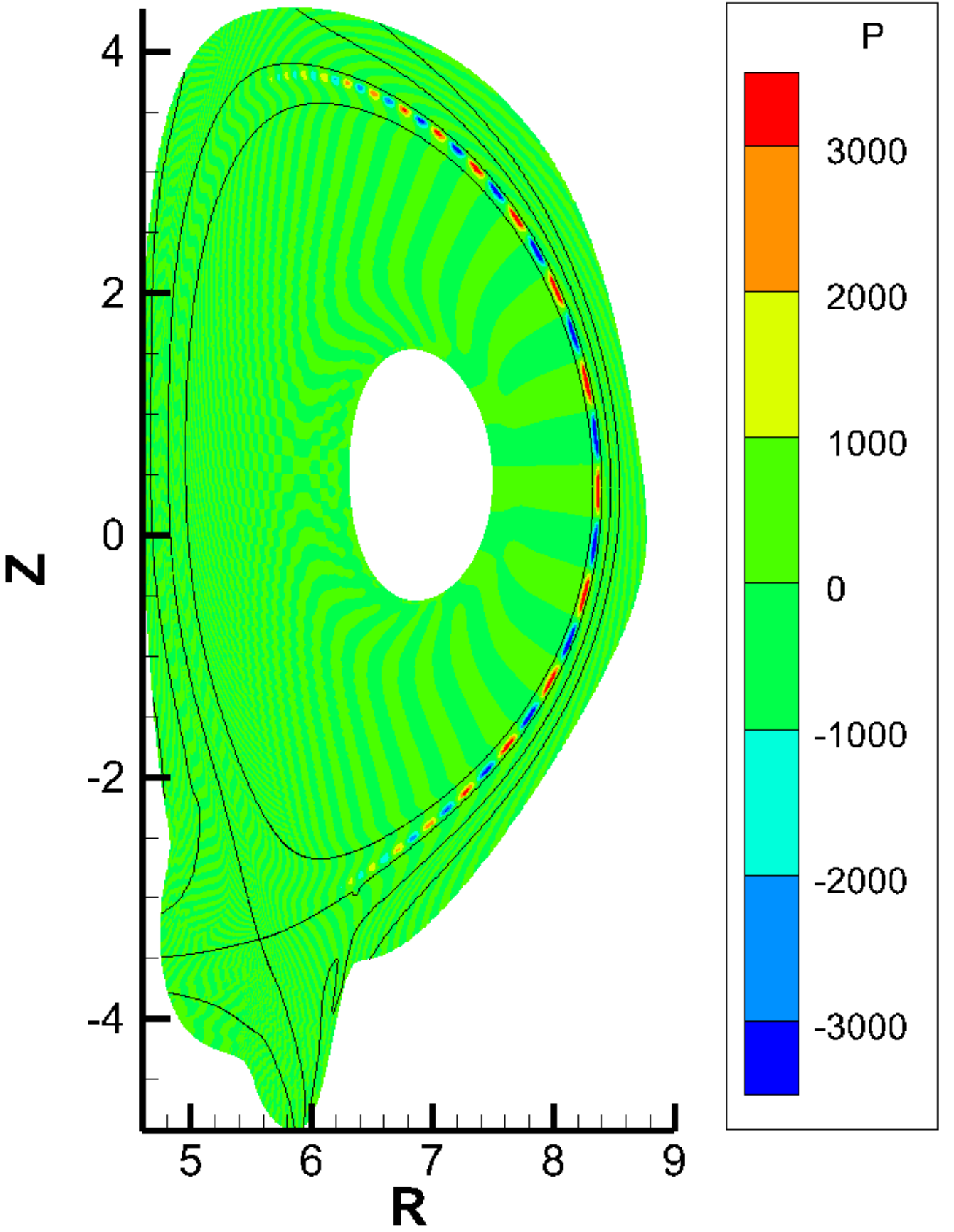}
\put(-220,200){\textbf{(c)}}
\put(-90,200){\textbf{n=10}}
\end{minipage}
\begin{minipage}{0.47\textwidth}
\includegraphics[width=1.0\textwidth,height=0.35\textheight]{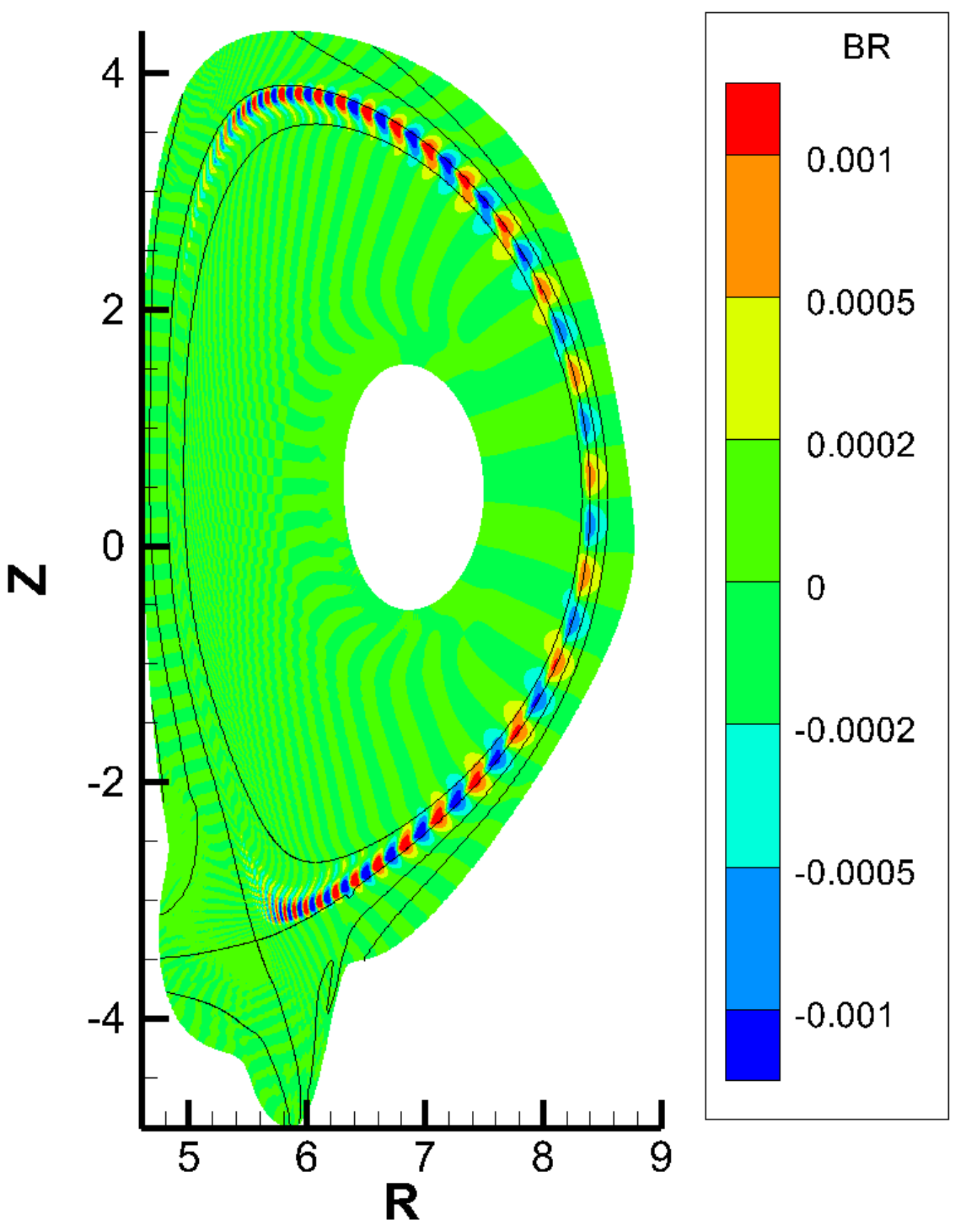}
\put(-220,200){\textbf{(d)}}
\put(-90,200){\textbf{n=10}}
\end{minipage}
\caption{Contour plots of CFETR proposed wall configuration in single-fluid MHD model. Colored-flood contour represent (a) pressure perturbation of $n=3$ mode, (b) radial magnetic perturbation of $n=3$ mode, (c) pressure perturbation of $n=10$ mode, (d) radial magnetic perturbation of $n=10$ mode. Solid-line contour represents equilibrium magnetic flux function.}
\label{fig:rw_contour}
\end{figure}

In addition, considering recently proposed wall of CFETR configuration, the growth rates of modes $n=1, 3, 5, 8, 10$  have also been carried out in single-fluid MHD model. These results are summarized in Fig.\ref{fig:rw_growth}. Three different self-similar wall positions with $b=1.2a, 1.4a$ and $1.6a$ are shown along with proposed wall shape. The growth rates of $n=1, 8, 10$ modes are almost same for the self-similar wall at $b=1.6a$, while those of $n=3, 5$ modes are very close to the self-similar wall at $b=1.4a$. The perturbed pressure and radial component of perturbed magnetic field contour plots of CFETR proposed wall configuration are shown in Fig.\ref{fig:rw_contour}, which indicates that wall configuration has little effect on the edge localized mode structure.

%=================================
% summary
%=================================

\section{Summary \& Discussions}
\label{summary}

In summary, stability of CFETR upgrade phase-I scenario has been studied in
context of ideal MHD analysis in single-fluid and two fluid MHD models. The initial-boundary value full MHD code NIMROD is employed to analyze the stability of $n=1-30$ modes numerically. It is predicted that all ideal MHD modes are unstable but edge localized in nature based on the single-fluid MHD model. No global core ideal MHD modes like internal kink mode is found to be dominantly unstable. In the two-fluid MHD analysis, all modes remain unstable and localized at the pedestal region. 
A clear yet weak effect of two-fluid stabilization on high-$n$ modes is noted, where all modes with $n>13$ have lower growth rates in two-fluid than in single-fluid MHD models. For modes with $n \leq 13$, two-fluid effects have little, if any, influence on the growth rates. Overall, the two-fluid stabilization is less than expected from the comentional local dispersion relation.

In addition, we have studied the wall position effects. The $n=1-10$ modes are found
to be more unstable with increasing wall position from plasma and finally the
growth rates of these modes approach the no wall limit value. On the other hand,
growth rates decrease as the wall position becomes closer to plasma boundary,  and all modes can be fully stabilized when the wall becomes sufficiently close. Taking the proposed wall of CFETR configuration into consideration, we have found the growth rates of $n=1, 3, 5, 8, 10$ modes are very close to those in the self-semilar wall cases at $b=1.4a$ and $b=1.6a$.

On basis of our analysis,  we conclude that the upgrade phase-I scenario of CFETR will not become dominantly unstable for
global ideal MHD modes. Such a design might help avoid disruption event
caused by ideal MHD instabilities. But, due to steep pedestal gradient and peaked edge
current, this scenario can suffer from medium to large size ELMs and the characteristics
of ELMs need to be determined from nonlinear simulation. The stable position of conducting wall is too close to plasma boundary to be a viable scheme for avoiding ELMs. To achieve long duration of
steady state operation maintaining fixed $\beta_N$, efficient schemes for ELM control are necessary. Among those schemes, the toroidal flow shear has been found influential on transforming large type-I ELMs
to grassy ELMs in experiments on present tokamaks. The toroidal rotation with self-consistent equilibrium pressure and density profiles
will be considered in our next evaluation of the linear stability of all toroidal modes. Finally, nonlinear 
simulation will be performed to quantify ELM frequency and heat flux to divertor plates and plasma facing components.

\begin{acknowledgements}
The research was supported by the National Magnetic Confinement Fusion Program of China under Grant Nos. 2014GB124002, 2015GB101004, and the 100 Talent Program
of the Chinese Academy of Sciences (CAS). Author D. B. is partially supported by CAS President International Fellowship Initiative (PIFI). Author P. Zhu also acknowledges the supports from U.S. DOE grant Nos. DE-FG02-86ER53218 and DE-FC02-
08ER54975. We are grateful for the support from the NIMROD team. This research used the computing resources from the Supercomputing Center of University of Science and Technology of China, 
and the National Energy Research Scientific Computing Center, a DOE Office of Science User Facility supported by the Office of Science of the U.S. Department of Energy under Contract No. DE-AC02-05CH11231.
\end{acknowledgements}
%\appendix
\newpage
\bibliography{cheng}
\newpage

%========================================
% Fig.1 cfetr grid
%========================================

\clearpage

%================================================
% Fig.2 equilibrium
%================================================

\clearpage
%========================================
% Fig.3 growth rate
%========================================

\clearpage
%========================================
% Fig.4 convergence
%========================================
%========================================
% Fig.5 contour plots
%========================================

\clearpage

%========================================
% Fig.6 contour plots
%========================================

\clearpage

%========================================
% Fig.7 wall position growth rate
%========================================

\clearpage

%========================================
% Fig.8 real wall growth
%========================================

\clearpage

%========================================
% Fig.9 real wall contour
%========================================

\clearpage

\end{document}